\documentstyle[12pt,a4]{article}
\begin{document}

\hbadness=10000
\hbadness=10000
\begin{titlepage}
\nopagebreak
\def\thefootnote{\fnsymbol{footnote}}
\begin{flushright}

        {\normalsize
DPSU-96-10\\
INS-Rep-1149\\
July, 1996   }\\
\end{flushright}
\vspace{1cm}
\begin{center}
\renewcommand{\thefootnote}{\fnsymbol{footnote}}
{\large \bf  Flat Directions in $Z_{2n}$ Orbifold Models}

\vspace{1cm}

{\bf Yoshiharu Kawamura $^a$ 
\footnote[1]{e-mail: ykawamu@gipac.shinshu-u.ac.jp}}
and  \ {\bf Tatsuo Kobayashi $^b$
\footnote[2]{e-mail: kobayast@ins.u-tokyo.ac.jp}}

\vspace{1cm}
$^a$ Department of Physics, Shinshu University \\

   Matsumoto 390, Japan \\
and\\
$^b$  Institute for Nuclear Study, University of Tokyo \\

       Midori-cho, Tanashi, Tokyo 188, Japan \\

\end{center}
\vspace{1cm}

\nopagebreak

\begin{abstract}
We study generic features related to matter contents and flat directions in
$Z_{2n}$ orbifold models.
It is shown that $Z_{2n}$ orbifold models have 
massless conjugate pairs, $R$ and $\overline R$,
in certain twisted sectors as well as one of untwisted subsectors.
Using these twisted sectors, $Z_{2n}$ orbifold models are 
classified into two types.
Conjugate pairs, $R$ and $\overline R$, lead to $D$-flatness as 
$\langle R \rangle = \langle \overline R \rangle \neq 0$.
We investigate generic superpotentials derived from orbifold 
models so as to show that this direction is indeed a flat direction.
\end{abstract}

\vfill
\end{titlepage}
\pagestyle{plain}
\newpage
\section{Introduction}
\renewcommand{\thefootnote}{\fnsymbol{footnote}}

Superstring theory is a promising candidate for the unified theory
of all interactions including gravity.
A number of 4-dimensional string models have been constructed
through several types of constructions, e.g.,
Calabi-Yau construction \cite{CY}, orbifold construction \cite{Orbi1}
and fermionic construction \cite{fermi}.
Although some interesting attempts have been done,
completely realistic models have not been found yet.

For example, it is shown that $Z_N$ orbifold models cannot have the standard 
model gauge group $SU(3) \times SU(2) \times U(1)$ without nontrivial 
Wilson lines \cite{Orbi3}.
The introduction of Wilson lines can lead to models
with smaller gauge groups including the standard model gauge group
and reduced matter multiplets
\cite{WL1}.
However, these models have, in general, several extra $U(1)$ symmetries 
and many extra matter fields besides those in the 
minimal supersymmetric standard model.
If the effective field theories derived from
string models have flat directions, gauge 
groups can break into smaller groups and extra matter fields can 
become massive.
Thus study on flat directions in those models is important 
from the viewpoint of realistic model constructions \cite{STflat}.
Actually some semirealistic models have been obtained through flat directions
in orbifold models \cite{STflat2}.
Further these flat directions could relate different models in string 
vacua.

Recently much interest is paid to understand classical and 
quantum moduli spaces (flat directions) of models with $N\geq 1$ 
supersymmetry (SUSY) \cite{SUSYdual} and string models 
with $N\geq 1$ and $d\geq 4$ \cite{STnp}.
Such a study leads to the understanding of
some nonperturbative aspects like dualities between 
SUSY models and string dualities.

In this way, study on flat directions is more and more important
in supersymmetric theories from several aspects. 
In the effective theories from $Z_N$ orbifold models,
flat directions have been investigated chiefly for $Z_3$ orbifold models 
as well as $Z_3 \times Z_3$ orbifold models
\cite{STflat,STflat2,STflat3}.
The other $Z_{2n}$ orbifold models have been little studied though
they have different features from $Z_3$ orbifold models.
For example, (2,2) $Z_{2n}$ orbifold models have $\overline {27}$ 
massless matter fields of the $E_6$ group as well as 27 massless matter 
fields, while the (2,2) $Z_3$ orbifold model as well as the (2,2) 
$Z_3 \times Z_3$ orbifold model has only 27 massless matter fields, 
but not $\overline {27}$ fields.
Thus in the $Z_{2n}$ orbifold models, we have candidates for flat 
directions as $\langle 27 \rangle =\langle \overline {27} \rangle \neq 0$.
In general, (0,2) $Z_{2n}$ orbifold models also have other 
conjugate pairs, $R$ and $\overline R$, and their VEVs could lead to a 
large symmetry breaking.
Such models should be paid attention to because 
they might lead to a realistic model through breakings by
flat directions.
If the models have generic features related to flat directions, 
we have to take into account them for a realistic model building
 from the beginning.

Conjugate massless pairs, $R$ and $\overline R$, appear not 
only in $Z_{2n}$ orbifold models but also in 4-dimensional string models
by other constructions, e.g. Calabi-Yau 
compactifications and fermionic constructions.
Thus study on flat directions, 
$\langle R \rangle=\langle \overline R \rangle \neq 0$, 
in $Z_{2n}$ orbifold models is also interesting as an example for 
generic four-dimensional string models.

In this paper, we study generic features related to matter contents and
flat directions in $Z_{2n}$ orbifold models.
It is shown that $Z_{2n}$ orbifold models have 
massless conjugate pairs, $R$ and $\overline R$,
in certain twisted sectors as well as one of untwisted subsectors.
Using these twisted sectors, $Z_{2n}$ orbifold models are 
classified into two types.
This classification is very useful to study flat directions
and the breaking mechanism by them in
$Z_{2n}$ orbifold models.
Conjugate pairs, $R$ and $\overline R$, lead to $D$-flatness as 
$\langle R \rangle = \langle \overline R \rangle \neq 0$.
We investigate generic superpotentials derived from orbifold 
models so as to show that vacuum expectation values (VEVs) of those 
pairs also lead generally to flat directions.

This paper is organized as follows.
In the next section, massless spectra in $Z_{2n}$ orbifold models 
are reviewed.
Also selection rules for couplings in orbifold models are reviewed.
In section 3, we study how conjugate pairs $R$ and $\overline R$ 
appear in massless spectra of orbifold models.
In subsection 3.1, it is shown that such pairs always appear in 
specific twisted and untwisted subsectors of $Z_{2n}$ orbifold models.
In terms of these sectors, we classify $E_8$ shifts to construct 
$Z_{2n}$ orbifold models into three classes in subsection 3.2, and 
further using combinations of $E_8$ shifts and $E'_8$ shifts and 
twisted sectors with conjugate pairs, $Z_{2n}$ orbifold models are
classified into only two types in subsection 3.3.
In subsection 4.1, it is shown VEVs of these conjugate pairs can lead 
to flat directions.
These are generic flat directions which all $Z_{2n}$ orbifold models 
have.
In order to show the existence of flat directions, we discuss selection
rules for renormalizable and nonrenormalizable couplings.
In subsection 4.2, we study an explicit model to ascertain the results
in subsection 4.1.
We show that explicit models have flat directions other than generic 
flat directions.
Also it is shown how many massless matter fields gain masses 
along flat directions.
In subsection 4.3, the situation of flat directions in models 
with nontrivial Wilson lines is discussed.
Wilson lines resolve degeneracy of massless matter fields, but 
similar flat directions can be found.
In subsection 4.4, comments related to anomalies are given.
Section 5 is devoted to conclusions and discussions.
In Appendix A, structures of $Z_{2n}$ orbifold models are summarized.

\section{Orbifold Models}
\subsection{Massless spectrum}

The $E_8 \times E'_8$ heterotic string theory consists of a bosonic 
string in the (4+6)-dimensional space-time, its right-moving superpartner 
and a left-moving gauge part, whose momenta $P^I$ ($I=1 \sim 16$) 
span an $E_8 \times E'_8$ lattice $\Gamma_{E_8 \times E'_8}$.
Nonzero $E_8$ root vectors are represented as 
\begin{eqnarray}
&(\underline {\pm 1, \pm 1,0,0,0,0,0,0}),
\label{vector} \\
&(\pm {1 \over 2},\pm {1 \over 2},\pm {1 \over 2},\pm {1 \over 2},
\pm {1 \over 2},\pm {1 \over 2},\pm {1 \over 2},\pm {1 \over 2}),
\label{spinor}
\end{eqnarray}
where the underline denotes any possible permutation of elements and 
the number of minus signs in Eq.~(\ref{spinor}) should be even.
It is obvious that if $P^I$ belongs to Eq.(\ref{vector}) or 
(\ref{spinor}), $-P^I$ also belongs.
The right-moving fermionic string is bosonized into 
the bosonic string whose momenta $p^t$ ($t=0 \sim 4$) 
span an $SO(10)$ weight lattice $\Gamma_{SO(10)}$.
Vector and spinor weights correspond to bosonic and 
fermionic fields, respectively.

In orbifold models \cite{Orbi1,Orbi2,Orbi3,Orbi4}, 
the 6-dimensional space is compactified 
into an orbifold, which is a division of a 6-dimensional 
torus $T^6$ by its automorphism (twist) $\theta$.
$Z_N$ orbifold models satisfy $\theta^N=1$ \footnote{
We have another type of orbifold models, which have 
two independent twists and are called 
$Z_N \times Z_M$ orbifold models \cite{ZNM}.}.
To preserve the world-sheet supersymmetry and the 
modular invariance, this twist $\theta$ 
should be associated with shifts on $\Gamma_{SO(10)}$ and 
$\Gamma_{E_8 \times E'_8}$, $v^t$ and $V^I$.
Here the elements corresponding to the 4-dimensional space-time, 
$v^t$ ($t=0,4$), vanish and the other elements are related with 
eigenvalues of $\theta$ as 
$\theta ={\rm diag} \exp [2 \pi iv^i]$ ($i=1,2,3$).
For the shift $V^I$, the consistency with $\theta^N=1$ leads 
to the constraint that $NV^I$ should be on 
$\Gamma_{E_8 \times E'_8}$.
This number $N$ is called the order of this shift $V^I$.
Further the modular invariance requires the following relation 
between $V^I$ and $v^i$:
\begin{eqnarray}
N\sum_{i=1}^3(v^i)^2-N\sum_{I=1}^{16}(V^I)^2={\rm even}.
\label{modinv}
\end{eqnarray}

The twists leading to the $N=1$ space-time supersymmetry in $d=4$ are 
classified into nine combinations of eigenvalues of $\theta$, i.e.,  
$Z_3$, $Z_4$, $Z_6$-I, $Z_6$-II, $Z_7$, $Z_8$-I, $Z_8$-II, 
$Z_{12}$-I and $Z_{12}$-II orbifold models.
The corresponding values of $v^i$ are shown in Appendix A except 
$Z_3$ and $Z_7$ orbifold models.
These twists can be realized as Coxeter elements on 
Lie lattices $\Gamma_6$, which are used to construct $T^6$ as
$T^6=R^6/\Gamma_6$.
The Coxeter element is a product of all Weyl reflections 
corresponding to the Lie lattice.
For example, the $Z_4$ orbifold has $v^i=1/4(1,1,-2)$.
The corresponding twist $\theta$ is obtained as the 
Coxeter element of an $SO(5)^2 \times SU(2)^2$ lattice.
We denote simple roots by $e_a$ ($a=1 \sim 6$), where 
one of the $SO(5)$ lattices is spanned by $(e_1, e_2)$ or 
$(e_3, e_4)$ and one of the $SU(2)$ lattices is spanned 
by $e_5$ or $e_6$.
These vectors are transformed under $\theta$ as 
\begin{eqnarray}
&\theta e_a=e_a+2e_{a+1}, \quad 
\theta e_{a+1}=-e_a-e_{a+1}, &\quad (a=1,3),
\label{Cox1} \\
&\theta e_a=-e_a, &\quad (a=5,6).
\label{Cox2}
\end{eqnarray}
It is easy to show this twist $\theta$ has eigenvalues 
$\exp [2 \pi i/4(1,1,-2)]$.
Coxeter elements of other lattices can realize 
these eigenvalues.
For example, the Coxeter elements of the 
$SO(5)\times SU(2) \times SU(4)$ and 
$SU(4)^2$ lattices have the same eigenvalues.
We can obtain the $Z_4$ orbifold as a division of the 
$SO(5)^2 \times SU(2)^2$ torus by this twist $\theta$ (\ref{Cox1}) 
and (\ref{Cox2}).
Similarly we can construct other $Z_N$ orbifolds.

There are two types of closed strings on orbifolds.
One is closed on tori and is called an untwisted string.
The other is a twisted string.
For the $\theta^k$-twisted sector $T_k$, the string 
coordinate satisfies the following boundary condition:
\begin{eqnarray}
x^i(\sigma=2\pi)=(\theta^k x)^i(\sigma=0)+e^i,
\label{bound}
\end{eqnarray}
where $e^i$ is a lattice vector.
Zero-modes of the twisted string satisfy the same condition 
as Eq.~(\ref{bound}).
These zero-modes are called fixed points and denoted by 
corresponding space group elements $(\theta^k,e^i)$.
For example, the $\theta$-twisted sector of the above $Z_4$ 
orbifold has 16 fixed points as 
\begin{eqnarray}
(\theta,ie_2+je_4+ke_5+\ell e_6),
\end{eqnarray}
where $i, j, k,\ell=0, 1$.

Further the $\theta^2$-twisted sector has 16 fixed points 
because each $SO(5)$ plane has four fixed points as 
\begin{eqnarray}
(\theta^2,ie_a+je_{a+1}),
\label{fix2}
\end{eqnarray}
where $i, j=0, 1$ and $a=1,3$.
Note that the $e_5-e_6$ plane is completely fixed under $\theta^2$.
Among Eq.~(\ref{fix2}), the fixed points $(\theta^2,ie_a)$ are also 
fixed under $\theta$.
However, the others are transformed under $\theta$ as 
\begin{eqnarray}
(\theta^2,e_{a+1}) \longleftrightarrow (\theta^2,e_a+e_{a+1}),
\end{eqnarray}
up to their conjugacy class.
Physical states should be eigenstates of $\theta$.
Hence we have to take linear combinations of the states 
corresponding to the above fixed points as \cite{Yukawa,Orbi4}
\begin{eqnarray}
|\theta^2,e_{a+1}\rangle 
\pm |\theta^2,e_a+e_{a+1} \rangle.
\end{eqnarray}
These states have eigenvalues $\pm 1$, respectively.
In addition, we have two eigenstates, 
$|\theta^2,ie_a \rangle$, whose eigenvalue is 1.
We construct ground states for the string 
on the 6-dimensional orbifold as tensor products of 
these states corresponding to two $SO(5)$ planes.
Then we have 10 states with the eigenvalue +1 and 
6 states with the eigenvalue $-1$, as 
explicitly shown in Appendix A.
The same structure on fixed points is obtained 
from the different lattices, $SO(5)\times SU(2) \times SU(4)$ 
and $SU(4)^2$.

In general, the $\theta^k$-twisted sector of the $Z_N$ orbifold model 
has the following $\theta$-eigenstates:
\begin{eqnarray}
|\theta^k,e^i \rangle +e^{-i\gamma}|\theta^k,\theta e^i \rangle +
\cdots + e^{-i\gamma(m-1)}|\theta^k,\theta^{m-1}e^i \rangle,
\label{state}
\end{eqnarray}
where $(\theta^k,e^i)$ is a fixed point of $T_k$  
and $m$ is the least number so that $(\theta^k,e^i)$ is 
fixed under $\theta^m$.
This state has an eigenvalue $e^{i\gamma}$ where $\gamma=2 \pi \ell/m$ 
with $\ell=$ integer.
The $T_n$ sectors of $Z_{2n}$ orbifold models have the 
same 16 fixed points as those of $T_2$ in $Z_4$ orbifold 
models, i.e., 
\begin{eqnarray}
(\theta^n,ie_1+je_2+ke_3+\ell e_4),
\label{fp}
\end{eqnarray}
where $i,j,k,\ell =0,1$.
For the $T_n$ sectors of $Z_{2n}$ orbifold models, $\theta$-eigenstates 
are written explicitly in Appendix A

For the $T_k$ sector, the $SO(10)$ and $E_8 \times E'_8$ momenta 
are shifted as $p^t+kv^t$ and $P^I+kV^I$.
Independent shifts $V^I$ on the $E_8$ lattice for each $Z_N$ orbifold are  
classified in Ref.~\cite{shift}.
It is convenient to describe mass formulae in the light-cone gauge, 
where $SO(10)$ momenta are reduced into transverse 
$SO(8)$ momenta $p^t+kv^t$ ($t=1 \sim 4$).
The massless conditions are written as 
\begin{eqnarray}
{1 \over 2} \sum_{t=1}^4 (p^t+kv^t)^2+N^{(k)}_R+
c_k-{1 \over 2}=0,
\label{Rmass}
\end{eqnarray}
for the right-moving $T_k$ sector, and 
\begin{eqnarray}
{1 \over 2} \sum_{I=1}^{16} (P^I+kV^I)^2+N^{(k)}_L
+N_{E_8 \times E'_8}+c_k-1=0,
\label{Lmass}
\end{eqnarray}
for the left-moving $T_k$ sector, 
where 
\begin{eqnarray}
c_k \equiv {1 \over 2} \sum_{i=1}^3\eta^i_{(k)}(1-\eta^i_{(k)}), 
\quad \eta^i_{(k)} \equiv |kv^i|-{\rm Int}|kv^i|,
\end{eqnarray}
and $N^{(k)}_L$ ($N^{(k)}_R$) denotes the left-moving (right-moving) 
oscillator number.
Here the $T_k$ sector with $k=0$ corresponds to the untwisted sector 
$U$.
In such a case, we should consider an additional contribution
from the 6-dimensional internal space momenta to the massless
conditions.
Further $N_{E_8 \times E'_8}$ denotes the oscillator number 
for the gauge part and this contributes only for the untwisted 
sector.
For the untwisted sector, bosonic fields have SO(8) momenta
$p^t=(\underline{1,0,0,0})$.
For twisted sectors, the SO(8) momenta satisfying 
Eq.~(\ref{Rmass}) are shown in Refs.~\cite{Yukawa,Orbi4}.
For example, $T_n$ massless bosonic states of $Z_{2n}$ orbifold models 
have $c_n=1/4$ and $p^t+nv^t=1/2(1,1,0,0)$ in the base where 
$nv^t=1/2(1,1,0,0)$ (mod $Z$) as Appendix A.
Eq.~(\ref{Lmass}) for the untwisted sector $(k=c_k=0)$ is satisfied 
by the states with $\sum_I(P^I)^2=2$, i.e., 
Eqs.~(\ref{vector}) and (\ref{spinor}) as well as the states with 
$N_{E_8 \times E'_8}=1$ and vanishing momenta.

Physical states should be invariant under the full $Z_N$ 
transformation, i.e., transformations of the $SO(8)$ and 
$E_8 \times E'_8$ lattices by shifts $v^t$ and $V^I$ and 
the $\theta$-twist on the 6-dimensional orbifold.
For the untwisted sector, this $Z_N$ invariance 
leads to the following constraint on their momenta:
\begin{eqnarray}
\sum_{I=1}^{16}P^IV^I-\sum_{i=1}^3p^iv^i={\rm integer}.
\label{ZNinv}
\end{eqnarray}
The untwisted states with $p^t=(0,0,0,1)$, 
i.e. $\sum P^IV^I=$integer, correspond to unbroken gauge bosons 
and   
the untwisted states with $p^t=(\underline {1,0,0},0)$ 
correspond to bosonic massless matter fields.
We denote these three untwisted subsectors corresponding to 
$(1,0,0,0)$, $(0,1,0,0)$ and $(0,0,1,0)$ by $U_1$, $U_2$ 
and $U_3$, respectively.

The $T_k$ state with momenta $p^t+kv^t$ and 
$P^I+kV^I$ has an eigenvalue for the full $Z_N$ 
transformation as 
\begin{eqnarray}
\Delta_k & = & O^{(k)}e^{i\gamma}\tilde \Delta_k,\\
\tilde \Delta_k & = & {\rm exp} 
[2 \pi i ({k \over 2}\ (\sum_i(v^i)^2-\sum_I(V^I)^2) 
\nonumber \\
& & +\sum_I(P^I+kV^I)V^I
-\sum_i(p^i+kv^i)v^i ],
\label{GSO'}
\end{eqnarray}
where $\gamma$ is the contribution from 
the ground state for the  6-dimensional orbifold like 
Eq.~(\ref{state}) and $\tilde \Delta_k$ is a contribution of 
the $SO(8)$ and $E_8 \times E'_8$ parts.
Further $O^{(k)}$ denotes a contribution from oscillators.
We call $\Delta_k$ the (generalized) GSO phase \cite{Orbi2,GSO}.
Physical states should have $\Delta_k=1$, that is, 
the 6-dimensional ground states with 
$e^{i\gamma}= (O^{(k)}\tilde \Delta_k)^{-1}$ are selected 
as physical states.

\subsection{Selection rules for couplings}

Here we review selection rules for couplings \cite{couple} 
including nonrenormalizable couplings 
\cite{nonre,STflat3} \footnote{See also Ref.\cite{fmass}.}.
For bosonic states in the non-oscillated $T_k$ sector, 
vertex operators are obtained in the $-1$-picture as 
\begin{eqnarray}
V_{-1}=e^{-\phi}e^{iKx}e^{i(p+kv)H}e^{i(P+kV)x}\sigma_{f\gamma},
\label{VB}
\end{eqnarray}
where $\phi$ and $H$ denote the bosonized superconformal ghost and 
the bosonized right-moving fermionic string,  
and $e^{iKx}$ and $e^{i(P+kV)x}$ are the 4-dimensional part and 
the gauge part.
Here $\sigma_{f\gamma}$ is the twisted field associated with the fixed 
point $f$ and the eigenvalue $e^{i \gamma}$.
Also vertex operators of fermionic states in the $-1/2$-picture are 
written as 
\begin{eqnarray}
V_{-1/2}=e^{-\phi/2}e^{iKx}e^{i(p+kv)_fH}e^{i(P+kV)x}\sigma_{f\gamma},
\label{VF}
\end{eqnarray}
where the momenta $(p+kv)^i_f$ for fermionic states are obtained by 
those for bosonic states as $(p+kv)^i_f=(p+kv)^i-1/2$.
These vertex operators are related by the supersymmetry.
For the untwisted sector, we do not need $\sigma_{f\gamma}$.
We can change the picture by the picture changing operator  
\cite{FMS}, which includes 
\begin{eqnarray}
e^{\phi}e^{-i\alpha_i H}\partial X_i,
\end{eqnarray}
where $\alpha_1$, $\alpha_2$ and $\alpha_3$ are 
(1,0,0), (0,1,0) and (0,0,1), and $\partial X_i$ is an oscillator 
for the $i$-th plane.
Note that if we change pictures, oscillators are included even in the 
vertex operators (\ref{VB}) and (\ref{VF}), which correspond to 
the massless states with $N^{(k)}_L=0$ in Eq.(\ref{Lmass}).
Similarly we can write vertex operators corresponding to the massless 
states with $N^{(k)}_L \neq 0$ in Eq.(\ref{Lmass}).

Let us study $V_FV_FV^\ell_B$ couplings, where $V_F$ and $V_B$ are 
vertex operators for fermionic and bosonic states, respectively.
These vertex operators consist of several parts as Eqs. (\ref{VB}) 
and (\ref{VF}).
Each of them provides selection rules for nonvanishing couplings.
In nonvanishing couplings, a product of space group elements 
$(\theta^k,e^i)$ corresponding to $f$ should be equivalent to 
(1,0) up to their conjugacy class and a product of $e^{i\gamma}$ 
should be unity.
For 16 fixed points in $T_n$ sectors of $Z_{2n}$ orbifold models 
(\ref{fp}), couplings are allowed if a product 
of the corresponding space group elements satisfies 
$\prod (\theta^n,i_ae_1+j_be_2+k_ce_3+\ell_d e_4)=(1,0)$, 
i.e..
\begin{eqnarray}
\sum_a i_a={\rm even}, \quad \sum_b j_b={\rm even}, \quad 
\sum_c k_c={\rm even}, \quad \sum_d \ell_d={\rm even}.
\label{space}
\end{eqnarray}
Further $SO(8)$ momenta as well as $E_8 \times E'_8$ momenta should 
be conserved.
In Refs.\cite{Yukawa,Orbi4}, renormalizable couplings satisfying 
$SO(8)$ momentum conservation are shown explicitly.
To match the background $\phi$-charge, the sum of $\phi$-charges for 
vertex operators should be $-2$.
Moreover nonvanishing couplings should be invariant under 
the following $Z_N$ transformation of oscillators;
\begin{eqnarray}
\partial X_{i(k)} \longrightarrow e^{2 \pi i kv^i}
\partial X_{i(k)},
\label{osci}
\end{eqnarray}
where $\partial X_{i(k)}$ denotes the oscillator of $T_k$ for the 
$i$-th plane.

For auxiliary fields of chiral fields, corresponding vertex 
operators in the 0-picture, $V_A$, are obtained from $V_{-1/2}$ 
through the supersymmetry \cite{ST-FI}.
Thus $V_AV_BV^\ell_B$ couplings have the same $\phi$-charge and 
the same $SO(8)$ momenta as $V_FV_FV^\ell_B$ couplings.
We can derive the same selection rules for $V_AV_BV^\ell_B$ 
couplings as $V_FV_FV^\ell_B$ couplings.

\section{Classification of $Z_{2n}$ orbifold models}

\subsection{Conjugate pairs in ${\hat U}_3$ and $T_n$ 
of $Z_{2n}$ orbifold models}

In this subsection, we show that
${\hat U}_3$ and $T_n$ sectors of $Z_{2n}$ orbifold models 
have pairs of massless fields with conjugate representations $R$ and 
$\overline R$ including massless fields with real representations.
Such pairs of fields are very important for study on flat directions 
because VEVs of these fields can lead to $D$-flatness.

Here ${\hat U}_3$ is a subsector in the untwisted sector with 
the momentum which satisfies $\sum_t p^t v^t = 1/2$.
The subsector ${\hat U}_3$ corresponds to the $U_3$ sector
in $Z_4$, $Z_6$-II, $Z_8$-II and $Z_{12}$-II orbifolds, 
which have $v^3=1/2$ (mod $Z$).
If $P^I$ satisfies $\sum_I P^I V^I=1/2$ (mod $Z$) for the $Z_N$ 
invariant condition (\ref{ZNinv}) in the ${\hat U}_3$ sector, 
$-P^I$ also satisfies $\sum_I P^I V^I=1/2$ (mod $Z$).
When $P^I$ corresponds to a particle with $R$ representation,
$-P^I$ corresponds to a particle with $\overline R$ representation.
Thus pairs of $R$ and $\overline R$ appear in massless 
spectra if one of them can appear.

Similar situations happen in the $T_n$ sector of $Z_{2n}$ orbifold models.
Suppose that a shifted momentum $P^I+nV^I$ satisfies the massless 
condition (\ref{Lmass}).
Then we always have the $E_8 \times E'_8$ momentum 
$P'^I=-P^I-2nV^I$ which sits on the $E_8 \times E'_8$ lattice 
because of $2nV^I=0$ (mod $\Gamma_{E_8 \times E'_8}$), 
and $P'^I+nV^I$ satisfies the same massless condition (\ref{Lmass}).
Thus conjugate pairs $R$ and $\overline R$ satisfy the massless 
condition (\ref{Lmass}) at the same time.

Next we study GSO phases for these conjugate pairs with 
$P^I+nV^I$ and $P'^I+nV^I$.
Since massless $T_n$ sectors have $p^t+nv^t=1/2(1,1,0,0)$ in $Z_{2n}$ 
orbifold models, 
we have 
\begin{eqnarray}
\delta \equiv 2n\sum_t (p^t+nv^t)v^t =1, 1, -1, -1, 2, -2 
{\rm \ and \ } 3,
\label{delta}
\end{eqnarray}
for $Z_4$, $Z_6$-I, $Z_6$-II, $Z_8$-I, $Z_8$-II, 
$Z_{12}$-I and $Z_{12}$-II orbifold models, respectively.
Further the modular invariance requires 
\begin{eqnarray}
\sum_t (v^t)^2-\sum_I(V^I)^2={m \over n},
\label{modular} 
\end{eqnarray}
where $m$ is an integer.
If $P^I+nV^I$ leads to ${\tilde \Delta_n} = (\omega_{(n)})^k$, i.e. 
\begin{eqnarray}
\sum_t (P^I+nV^I) V^I = {\delta \over 2n} 
- {m \over 2} + {k \over n} \quad {\rm mod \ }Z,
\label{GSO1}
\end{eqnarray}
where $\omega_{(n)} \equiv {\rm exp}(2\pi i/n)$ and $k = 0, 1,...,n-1$,
then $P'^I+nV^I$ satisfies 
\begin{eqnarray}
\sum_t (P'^I+nV^I) V^I = -{\delta \over 2n} 
+  {m \over 2} - {k \over n} \quad {\rm mod \ }Z.
\label{GSO2}
\end{eqnarray}
Hence the charge conjugation $(P+nV) \rightarrow -(P+nV)$ 
transforms $\tilde \Delta_n$ as \footnote{
This charge conjugation is similar to the CP transformation 
defined for orbifold models in Ref.\cite{CP}.
In this CP transformation the 6-dimensional compact space and its 
supersymmetric space are transformed simultaneously through their 
parity reflections, so that $\tilde \Delta_n$ transforms into 
$1/ \tilde \Delta_n$ under CP
and $\Delta_n$ is invariant.}
\begin{eqnarray}
{\tilde \Delta_n}=\omega_{(n)}^k \longrightarrow 
{\tilde \Delta_n} = (\omega_{(n)})^{-k - \delta}.
\label{conj}
\end{eqnarray}
Therefore pairs of $R$ and $\overline R$ appear in massless 
spectra of the $T_n$ sector if one of them can appear.\footnote{
As an exception, conjugate spinor representations such as $SO(16)$ 
do not appear because the adjoint representation in $E_8$ 
does not include such conjugate spinor representations.
However representations such as the $SO(16)$ spinor include 
both of $P+nV$ and $-(P+nV)$ for Eqs. (\ref{vector}) and (\ref{spinor}), 
as we shall show later.
}
The second column of Table 1 shows explicitly this transformation 
of $\tilde \Delta_n$ (\ref{conj}).

Note that if these $R$ and $\overline R$ fields 
correspond to different values of $\tilde \Delta_n$, 
their degeneracies are, in general, different from each other.
For example, we have 10 massless states with $R$ and $\Delta=1$, 
and 6 states with its conjugate representation 
$\overline R$ and $\Delta=-1$ in the ${T_2}$ sector 
of $Z_4$ orbifold models.
The third column of Table 1 shows the common number of 
degeneracy factors for ${\tilde \Delta_n}=\omega_{(n)}^k$ and 
${\tilde \Delta_n} = (\omega_{(n)})^{-k - \delta}$.

Moreover it is notable that massless fields with real representations 
such as 56 of $E_7$ 
can appear only if the value of $\tilde \Delta_n$ for $P^I+nV^I$ is 
same as one for $P'^I+nV^I$, i.e., 
$(\omega_{(n)})^k=(\omega_{(n)})^{-k-\delta}$.
Because all fields in a multiplet should have the same value for 
$\tilde \Delta_n$ and a real representation includes both of 
$P^I+nV^I$ and $P'^I+nV^I$.
For example, the $T_2$ sector of $Z_4$ orbifold models cannot 
have a real 56 representation of $E_7$ as a massless field.
Such real representations can appear in $T_3$ of $Z_6$-I and II, 
 $T_4$ of $Z_8$-II and $T_6$ of $Z_{12}$-I.

\subsection{Classification of shift vectors}

In the previous subsection, it is shown that $T_n$ sectors of 
$Z_{2n}$ orbifold models as well as the $\hat U_3$ sector 
can have conjugate pairs, $R$ and $\overline R$.
It is useful to classify shifts $V^I$ in terms of 
$V^I_{(n)}\equiv nV^I$ in 
order to classify $T_{n}$ sectors of $Z_{2n}$ orbifold models.
For any shift, $V^I_{(n)}$ satisfies $2V^I_{(n)}=0$ 
(mod $\Gamma_{E_8 \times E_8}$).
Thus $V^I_{(n)}$ should be a shift with the order 2.
There are only three independent shifts with the order 2 for 
each $E_8$ lattice.
These three shifts with the order 2, $V^I_2$, are classified 
in terms of $\sum_I (V^I_2)^2=0,1/2$ or 1, where $I$ runs from 1 to 8 or 
from 9 to 16.
These shifts $V^I_2$ are written as 
\begin{eqnarray}
&(0,0,0,0,0,0,0,0), 
\label{shift21} \\
&{1 \over 2}(1,1,0,0,0,0,0,0), 
\label{shift22} \\
& (1,0,0,0,0,0,0,0),
\label{shift23}
\end{eqnarray}
up to $E_8$ rotations.
The first shift does not break the $E_8$ group, while 
the others, Eqs.~(\ref{shift22}) and (\ref{shift23}) break 
the $E_8$ group into $E_7 \times SU(2)$ and $SO(16)$, respectively.

Here we classify shifts $V^I$ for $E_8$ in terms of corresponding 
$V^I_{(n)}$ into three classes, Classes 1, 2 and 3.
Shifts $V^I$ in Classes 1, 2 and 3 have $V^I_{(n)}$ which are 
equivalent to Eqs.~(\ref{shift21}), (\ref{shift22}) and 
(\ref{shift23}), respectively.
All independent shifts of $Z_N$ orbifold models for $E_8$ are obtained 
in Ref.~\cite{shift}.
Here we explicitly classify $E_8$ shifts for $Z_{4}$ and $Z_{6}$ orbifold 
models \cite{Z46} into three classes.

First we discuss $Z_4$ orbifold models, whose independent $E_8$ shifts 
are shown in the second column of Table 2.
Through the above classification, shifts $V^I$ 
are classified by $2V^I$ as follows.

Class 1: $2V=(0,\cdots,0)$ mod the $E_8$ lattice.
This class includes the following three shifts;
\begin{eqnarray}
&~&\#0\quad V=(0,\cdots, 0),  \quad V^2=0,\\
&~&\#1\quad V=(2,2,0,\cdots, 0)/4,  \quad  V^2=1/2,\\
&~&\#4\quad V=(1,0,\cdots, 0),  \quad  V^2=1,
\end{eqnarray}
where the number of the shift corresponds to that in Table 2.
This class of the shifts leads to no massless matter 
field in $U_1$ and $U_2$ sectors $(\sum_I P^I V^I=1/4)$.

Class 2: $2V=(1,1,0,\cdots,0)/2$ or $2V=(1,-1,0,\cdots,0)/2$ 
mod the $E_8$ lattice.
\begin{eqnarray}
&~&\#2\quad V=(1,1,0,\cdots, 0)/4,  \quad  V^2=1/8,\\
&~&\#3\quad V=(2,1,1,0,\cdots, 0)/4,  \quad V^2=3/8,\\
&~&\#6\quad V=(3,1,0,\cdots, 0)/4,  \quad V^2=5/8,\\
&~&\#8\quad V=(3,1,\cdots,1,0,0)/4,  \quad V^2=7/8.
\end{eqnarray}
Note that $2V=(1,1,0,\cdots,0)/2$ and $2V=(1,-1,0,\cdots,0)/2$ are 
equivalent $Z_2$ division of $E_8$.
This class of the shifts leads to 56 massless states satisfying 
$\sum_I P^I V^I=1/4$ in each of $U_1$ and $U_2$.

Class 3: $2V=(1,0,\cdots,0)$ mod the $E_8$ lattice.
\begin{eqnarray}
&~&\#5\quad V=(2,0,\cdots, 0)/4,  \quad V^2=1/4,\\
&~&\#7\quad V=(2,2,2,0,\cdots, 0)/4,  \quad  V^2=3/4,\\
&~&\#9\quad V=(1,\cdots,1,-1 )/4,  \quad  V^2=1/2.
\end{eqnarray}
This class of the shifts leads to 64 massless states 
satisfying $\sum_I P^I V^I=1/4$ in each of $U_1$ and $U_2$.

We can show the reason why shifts in one  
class lead to the same number of the $U_1$ and $U_2$ massless fields 
$(\sum_I P^I V^I=1/4)$ as follows.
For each shift we can write $2V^I$ as  
\begin{eqnarray}
2V^I=V_2^I+P_V^I,
\end{eqnarray}
where $V^I_2$ is one of Eqs.~(\ref{shift21}), (\ref{shift22}) and 
(\ref{shift23}), and $P_V^I$ sits on the $E_8$ lattice.
Now we discuss the following equation:
\begin{eqnarray}
\sum_IP^I(2V)^I={1 \over 2}, \quad ({\rm mod \ Z}).
\label{Z4U1}
\end{eqnarray}
Because of $\sum_I P^I P_V^I=$ integer, the number of the $E_8$ roots 
$P^I$ satisfying Eq.~(\ref{Z4U1}) depends only on $\sum_I P^I V_2^I$, 
but not on $\sum_I P^I P_V^I$.
Thus all of the shifts in one class give the same number 
of $P^I$ satisfying Eq.~(\ref{Z4U1}).
These $E_8$ roots $P^I$ satisfy $\sum_IP^IV^I=\pm 1/4$ (mod $Z$).
It is obvious that if $\sum_IP^IV^I=1/4$ (mod $Z$), its 
``conjugate'' momentum $(-P^I)$ satisfies $\sum_I(-P^I)V^I=-1/4$ 
(mod $Z$).
Hence the number of $P^I$ satisfying $\sum_IP^IV^I=1/4$ (mod $Z$) 
is a half of the number of $P^I$ satisfying Eq.~(\ref{Z4U1}).
Therefore all of the shifts in a class have the same number of 
massless $U_i$ matter fields for $i=1,2$.

In the same way, $E_8$ shifts are classified into 
the following three classes for $Z_6$ orbifold models.
The shifts, gauge groups and untwisted sectors are 
given in Table 3 explicitly.

Class 1: $3V=(0,\cdots,0)$ mod the $E_8$ lattice.
This class includes the following five shifts;
\#0, \#3, \#4, \#9 and \#17 in Table~3.
This class of the shifts leads to no massless matter 
field in the sector with $\sum_I P^I V^I=1/6$ or $3/6$.
In this case, shifts are written as 
$V^I=P^I_V/3$, where $P^I_V$ sits on the $E_8$ lattice.
Thus the inner product $\sum_IP^IV^I$ should satisfy 
$\sum_IP^IV^I=m/3$, where $m$ is an integer.
That is the reason why this class of shifts have no 
matter fields with $\sum_I P^I V^I=1/6$ or $3/6$.

Class 2: $3V=(1,1,0,...,0)/2 $ mod the $E_8$ lattice.
This class includes the following shifts;
\#1, \#2, \#5, \#6, \#11, \#12, \#14, \#15, \#20, \#22, \#23, and \#24 
in Table~3.

Class 3: $3V=(1,0,...,0) $ mod the $E_8$ lattice.
This class includes the following shifts;
\#7, \#8, \#10, \#13, \#16, \#18, \#19, \#21, \#25 and \#26 in Table~3.

In the same way, $E_8$ shifts of $Z_{8}$ and $Z_{12}$
orbifold models \cite{Z812} \footnote{See also 
Ref.\cite{Z8}.}
can be classified into three classes.
In general, shifts of Classes 2 and 3 break the $E_8$ into 
subgroups of $E_7 \times SU(2)$ and $SO(16)$, respectively.

\subsection{Classification of $T_{n}$ sector}

In the previous subsection, shifts for $E_8$ are classified into three 
classes.
Here we use this classification to study $T_n$ sectors in $Z_{2n}$ 
orbifold models.
Combinations of $E_8$ and $E'_8$ shifts are constrained due to 
the modular invariance Eq.~(\ref{modinv}).
As discussed in the previous subsection, both of $E_8$ and $E'_8$ 
shifts for $Z_{2n}$ orbifold models, $V^J$ ($J=1 \sim 8$) and 
$V^K$ ($K=9 \sim 16$), are written as 
\begin{eqnarray}
nV^J=V^J_2+P^J_V, \quad nV^K=V^K_2+P^K_V,
\end{eqnarray}
where $P^I_V$ sits on $\Gamma_{E_8 \times E'_8}$ and $V^I_2$ is obtained as 
Eqs.~(\ref{shift21}), (\ref{shift22}) and (\ref{shift23}).
They satisfy 
\begin{eqnarray}
\sum_{J=1}^8(V^J_2)^2={m_{E_8} \over 2}, \quad 
\sum_{K=9}^{16}(V^K_2)^2={m_{E'_8} \over 2},
\end{eqnarray}
where $m_{E_8}, m_{E'_8}=0,1$ and 2 for Classes 1, 2 and 3, respectively.
For $T_n$ sectors of $Z_{N=2n}$ orbifold models, we obtain 
the following equation;
\begin{eqnarray}
N\sum_{I=1}^{16}(V^I)^2={m_{E_8} + m_{E'_8} +2\ell \over n},
\end{eqnarray}
where $\ell=$ integer, because $2V^I_2$ sits on the $E_8$ or 
$E'_8$ lattice, i.e. $2\sum_IV^I_2P^I_V=$ integer.
On the other hand, we use explicit values of $v^i$ to find 
\begin{eqnarray}
nN\sum_i (v^i)^2={\rm odd},
\end{eqnarray}
for all $Z_{N=2n}$ orbifold models.
Thus the modular invariance (\ref{modinv}) requires 
$m_{E_8} + m_{E'_8}$ should be odd and allows the combinations where 
only one of the $E_8$ shift and the $E'_8$ shift 
belongs to Class 2, i.e. combinations of Classes 2 \& 1 and 3 \& 2.
The other four combinations are forbidden.
Note that the $Z_{2n}$ orbifold models with the unbroken 
$E_8 \times E'_8$ gauge group are not allowed, while the 
$Z_3$ orbifold model with the unbroken 
$E_8 \times E'_8$ gauge group are allowed.

Then we have only two types of $T_n$ massless spectra of $Z_{2n}$ 
orbifold models.
In one type (Classes 2 \& 1), the $T_n$ massless condition for 
$Z_{2n}$ orbifold models (\ref{Lmass}) is 
satisfied with 28 momenta $P+nV$ for $N^{(n)}_L=0$ as 
\begin{eqnarray}
& (- 1/2,- 1/2,\pm1,0,\cdots,0)(0,\cdots,0),\nonumber \\
& (1/2,1/2,0,\underline {\pm 1,0,0,0,0})(0,\cdots,0),
\label{T11}\\
& (0,0,1/2,\pm 1/2,\cdots,\pm 1/2)(0,\cdots,0),
\nonumber
\end{eqnarray}
and their ``conjugates'' $-(P+nV)$, and two momenta $(P+nV)$ 
for $N^{(n)}_L=1/2$ as
\begin{eqnarray}
(\underline {1/2,-1/2},0,\cdots,0)(0,\cdots,0),
\label{T12}
\end{eqnarray}
in the base where $nV=1/2(1,-1,0,\cdots,0)(0,\cdots,0)$ 
mod the $E_8 \times E_8$ lattice.
Totally the 56(=28+28) states correspond to 
a 56 representation of $E_7$ if it is unbroken.
And two states with $N^{(n)}_L=1/2$ correspond to a doublet representation 
of $SU(2)$.
In general, this type has a gauge group 
$\tilde G(E_7) \times \tilde G(SU(2)) \times \tilde G(E'_8)$, 
where $\tilde G(G)$ denotes a subgroup of $G$.
Thus $T_n$ sectors of $Z_{2n}$ orbifold models do not have 
massless matter fields with any nontrivial representations 
under $\tilde G(E'_8)$.

For example, the modular invariance allows twelve $Z_4$ orbifold 
models as shown in Table 4, where the third column shows combinations 
of $E_8$ shifts and $E'_8$ shifts as the corresponding numbers of the 
first column in Table 2.
Massless states of $T_1$ and $T_2$ are shown in 
the fourth and fifth columns of Table 4.
Models of No. 1, 2, 3, 8, 9 and 10 in Table 4 correspond to 
combinations of shifts Classes 2 \& 1. 
Actually all of these $T_2$ sectors have the following massless states;
\begin{eqnarray}
N^{(2)}_L=0, \quad {\tilde \Delta_2}=1, \quad 10 \times 28,\\
N^{(2)}_L=0, \quad {\tilde \Delta_2}=-1, \quad 6 \times 28,\\
N^{(2)}_L=1/2, \quad {\tilde \Delta_2}=1, \quad 2 \times 10 \times 2,\\
N^{(2)}_L=1/2, \quad {\tilde \Delta_2}=-1, \quad 2 \times 6 \times 2,
\end{eqnarray}
where the degeneracy factor 10 (6) for ${\tilde \Delta_2}=1 (-1)$ 
is generated by fixed points and the degeneracy 
factor 2 for $N^{(2)}_L=1/2$ is generated by two oscillators 
corresponding to the first and second planes, 
$N^{(2)}_{L1}$ and $N^{(2)}_{L2}$.
These 28 states correspond to (27 + singlet) in unbroken 
$E_6$ or 28 antisymmetric representation in unbroken $SU(8)$.

In the other type (Classes 3 \& 2), the massless condition for $T_n$ of 
$Z_{2n}$ orbifold models (\ref{Lmass}) is 
satisfied with 16 momenta $P+nV$ for $N^{(n)}_L=0$ as 
\begin{eqnarray}
(\underline {\pm 1,0,\cdots,0})(1/2,-1/2,0,\cdots,0),
\label{T2}
\end{eqnarray}
and their ``conjugates'' $-(P+nV)$ 
in the base where 
\begin{eqnarray}
nV=(1,0,\cdots,0)(1/2,-1/2,0,\cdots,0) \quad 
({\rm mod}\ \Gamma_{E_8 \times E'_8}).
\end{eqnarray}
These 32(=16+16) states correspond to $(16_v,2)$ under 
$SO(16) \times SU(2)'$ if it is not broken.
There is no massless matter fields with nonvanishing $N^{(n)}_L$.
In general, this type has a gauge group 
$\tilde G(SO(16)) \times \tilde G(SU(2)') \times \tilde G(E'_7)$, 
The above momenta do not have any quantum numbers under 
$\tilde G(E'_7)$.

For example, models of No. 4, 5, 6, 7, 11 and 12 in Table 4 
of $Z_4$ orbifold models correspond to this type.
These $T_2$ sectors have the following massless spectrum;
\begin{eqnarray}
N^{(2)}_L=0, \quad {\tilde \Delta_2}=1, \quad 10 \times 16,\\
N^{(2)}_L=0, \quad {\tilde \Delta_2}=-1, \quad 6 \times 16.
\end{eqnarray}
These 16 states correspond to 14 + 1 + 1 for unbroken 
$SO(14)$, (10,1) and (1,6) for unbroken $SO(10) \times SU(4)$ 
or (8,2) for unbroken $SU(8) \times SU(2)'$.

Taking into account the unbroken gauge groups, 
we can classify twelve $Z_4$ orbifold models of Table 4 as 
\begin{eqnarray}
&~&1) \quad E_6 \times SU(2) \times U(1) \times \tilde G(E'_8) 
{\rm \ models},
\label{model1} \\
&~&2) \quad SU(8) \times SU(2) \times \tilde G(E'_8) {\rm \ models},\\
&~&3) \quad SO(14) \times U(1) \times U(1)' \times \tilde G(E'_7) 
{\rm \ models},\\
&~&4) \quad SO(10) \times SU(4) \times U(1)' \times \tilde G(E'_7) 
{\rm \ models},\\
&~&5) \quad SU(8) \times U(1) \times SU(2)' \times \tilde G(E'_7) 
{\rm \ models},
\end{eqnarray}
where the first and second types of models correspond to 
the combination Classes 2 \& 1 and the other models correspond 
to the combination Classes 3 \& 2.
The latter type of models include anomalous $U(1)$ 
symmetries.\footnote{This fact might provides a condition that an 
anomalous $U(1)$ symmetry appears.
Some conditions for absence of anomalous $U(1)$ are shown in 
Ref.\cite{KN}.}

In the same way, we can classify $T_n$ sectors of $Z_{2n}$ 
orbifold models into two types, combinations of 
Classes 2 \& 1 and 3 \& 2.
That is very important.
There are many independent $Z_N$ orbifold models, 
e.g., 58 $Z_6$-I, 61 $Z_6$-II, 246 $Z_8$-I, 248 $Z_8$-II, 3026 $Z_{12}$-I 
and 3013 $Z_{12}$-II orbifold models \cite{Orbi3}, 
but these are classified into only the above two types by 
the structure of $T_n$ sectors.
For example, $T_3$ sectors of all $Z_6$-I and II orbifold models 
are shown in Table 5 and 6, where 
$G_{6,3,2}$ denotes $SU(6)\times SU(3) \times SU(2)$.
The third columns of Table 5 and 6 show combinations of $E_8$ shifts 
and $E'_8$ shifts as the corresponding numbers of the first column in 
Table 3.
The combinations of Classes 2 \& 1 correspond to the 
following $Z_6$-I orbifold models;
\begin{eqnarray}
{\rm No.}1,2,3,4,9,10,12,13,17,19,20, \nonumber \\
26,27,33,39,41,44,45 {\rm \ and \ } 49,
\end{eqnarray}
and the following $Z_6$-II orbifold models;
\begin{eqnarray}
{\rm No.}1,2,3,4,5,6,11,12,15,20,21,25, \nonumber \\
26,27,34,35,41,44,46,49 {\rm \ and \ } 50.
\end{eqnarray}
The other $Z_6$ orbifold models correspond to Classes 3 \& 2.
Every massless state corresponds to Eq.(\ref{T11}), (\ref{T12}),  
(\ref{T2}) or their ``conjugates''.
However, degeneracy factors are not so simple as those of 
$T_2$ of $Z_4$ orbifold models.
Because the $T_3$ sectors of $Z_6$-I and II orbifold models have 
three values for $e^{i\gamma}$ (\ref{state}), i.e. 1, 
$\omega$ and $\omega^2$, 
where $\omega$ is the third root of unity, and these values of 
$e^{i\gamma}$ in $Z_6$-I (II) orbifold models correspond to 
degeneracy factors, 6, 5 and 5 (8, 4 and 4), respectively.
Further, in general, each multiplet can have a different value of 
$\tilde \Delta_{(n)}$ from others.
For example, in the $Z_6$-I orbifold model with No. 52, 
($6,1,1;1,2$) and its ``conjugate'' ($\overline 6,1,1;1,2$) have 
$\tilde \Delta_{3}=1$ and $\omega^2$, leading to degeneracy 
factors 6 and 5, respectively.
This relation of $\tilde \Delta_{3}$ is consistent with Eq.(\ref{conj}).
On the other hand, ($1,2,2;1,2$) has $\tilde \Delta_{3}=\omega$ and 
degeneracy factor 5.
However, in any case we can find at least five pairs of $P+3V$ 
and $-(P+3V)$ as shown in Table 1.
Similarly we can find at least four conjugate pairs in the $T_3$ sector 
of $Z_6$-II orbifold models making use of the relation (\ref{conj}) 
and corresponding degeneracy factors.
Further $T_4$ of $Z_8$-I and II  orbifold models and 
$T_6$ of $Z_{12}$-I and II orbifold models have at least three and 
two conjugate pairs, respectively.
Moreover we find at least four (three) conjugate pairs in 
$T_4$ ($T_6$) of $Z_8$-II ($Z_{12}$-I) orbifold models if 
corresponding $P+nV$ or $-(P+nV)$ leads to $\tilde \Delta_n=1$ 
($\omega^m$).
These common numbers of degeneracy factors for conjugate pairs 
are shown in the third column of Table 1.
These numbers are very important for the study of gauge
symmetry breaking by flat directions.

\section{Flat directions}
\subsection{Generic flat directions in $T_n$ and $\hat U_3$ 
sectors of $Z_{2n}$ orbifold models}

Some conjugate pairs of $T_n$ matter fields with momenta 
$P+nV$ and $-(P+nV)$, $T_{n,P+nV}$ and $T_{n,-(P+nV)}$,  
can appear in $Z_{2n}$ orbifold models including 
conjugate pairs of the $\hat U_3$ sector .
VEVs of these fields are important for the study on flat directions, 
because they lead to the following $D$-flat direction;
\begin{eqnarray}
\langle T_{n,P+nV}\rangle =\langle T_{n,-(P+nV)} \rangle \neq 0.
\label{Dflat}
\end{eqnarray}
If this direction (\ref{Dflat}) is also a flat direction for $F$-terms, 
a rank of a gauge group is reduced by a degeneracy of these 
conjugate pairs.
In this subsection, we study whether this direction (\ref{Dflat}) 
is really a flat direction for the superpotential $W$ derived 
from orbifold models.
Flat directions mean 
\begin{eqnarray}
\langle W \rangle =\langle {\partial W \over \partial \chi }\rangle 
=0,
\end{eqnarray}
as well as vanishing $D$-terms where $\chi$ is any chiral
superfield.

Let us consider flat directions (\ref{Dflat}) in $T_n$ sectors of 
$Z_{2n}$ orbifold models.
In this case, we have to examine $(T_n)^\ell$ and 
$\chi(T_n)^\ell$ couplings.
Here we restrict ourselves to renormalizable couplings in $W$ 
for a while.
In this case, the point group selection rule allows 
only $(T_n)^2$ couplings and $U_i(T_n)^2$ couplings among couplings 
relevant to flat directions (\ref{Dflat}) of $T_n$.
The $T_n$ sectors have the $SO(6)$ momentum $p+nv=(1,1,0)/2$ and 
the total momentum of $U_i(T_n)^2$ couplings is reduced 
by $(1,1,1)$ through twice 
supertransformation to obtain the $V_FV_FV_B$ or $V_AV_BV_B$ form.
Thus only the $U_3(T_n)^2$ coupling is allowed because of the $SO(6)$ 
momentum conservation.
Further a product of $e^{i\gamma}$ for coupled states should be 
unity.
That means a product of corresponding $\tilde \Delta_n$ 
should be unity, if their coupling is allowed.
Therefore the conjugate pairs, $R$ and $\overline R$, cannot 
couple, because a product of the corresponding values 
for $\tilde \Delta_n$ is obtained as 
$\omega^k_{(n)}\omega^{-k-\delta}_{(n)}=
\omega^{-\delta}_{(n)}$ due to the relation (\ref{conj}), and 
$\delta/n \neq$ integer (\ref{delta}).
Moreover the space group selection rule requires that 
coupled $T_n$ states sit the same fixed points 
because of Eq.(\ref{space}).

For example, the $Z_4$ orbifold models have the following 
superpotential;
\begin{eqnarray}
W& =& \sum_a U_{3,P_U}T_{n,a,P_1+nV}T_{n,a,P_2+nV} 
\nonumber \\
& + & \sum_b U_{3,P_U}T_{n,b,-(P_1+nV)}
T_{n,b,-(P_2+nV)},
\label{poten}
\end{eqnarray}
where $a$ ($b$) denotes 10 (6) states with $e^{i\gamma}=1$ ($-1$) and 
coupling strengths are omitted.
Note that in $W$ the same 6-dimensional states are allowed to couple.
Thus Eq.(\ref{Dflat}) is a flat direction if there does not 
exist the $U_3$ field satisfying $P_U\pm (P_1+nV) \pm (P_2+nV)=0$ (i.e. 
gauge invariance) in massless spectra.
Here we make six pairs of conjugate states with 
$\tilde \Delta_2=1$ and $-1$ and then consider the following VEVs;
\begin{eqnarray}
\langle T_{n,a=c,P_c+nV}\rangle = 
\langle T_{n,b=c,-(P_c+nV)}\rangle \neq 0,
\label{flat}
\end{eqnarray}
where $c=1 \sim 6$ and $T_{n,a=c,P_c+nV}$ are six of ten 
degenerate states with $e^{i \gamma}=1$.
Eq.(\ref{flat}) implies that for each ``fixed point'' $c$ only one 
field (in a multiplet) with $P_c+nV$ develops its VEV.
In this case, couplings relevant to flat directions are the couplings 
(\ref{poten}) with $P_1+nV=P_2+nV$.
This momentum is one of Eqs.(\ref{T11}), (\ref{T2}) and their 
``conjugates''.
Among Eqs. (\ref{T11}), (\ref{T2}) and their ``conjugates'', 
there is no momentum for $T_n$ 
satisfying $P_U+2(P+nV)=0$ with the $E_8$ root vector $P_U$, 
i.e. (\ref{vector}) and (\ref{spinor}).
Thus there is no $U_3$ massless matter field which have a 
coupling with $(T_{n,a=c,P+nV})^2$ or $(T_{n,b=c,-(P+nV)})^2$, 
due to the $E_8$ momentum conservation.
That implies that 
the group $\tilde G(E_7)$ ($\tilde G(SO(16)$) could break along 
this flat direction (\ref{flat}) reducing its rank by at least 6
in the type of models with Classes 2 \& 1 (3 \& 2) shifts.
Also $\tilde G(SU(2)')$ in the Classes 3 \& 2 type of models could 
break.
Further $\tilde G(SU(2))$ in the Classes 2 \& 1 type of models could break 
through VEVs of the massless matter fields with the momenta (\ref{T12}) 
and $N^{(2)}_L=1/2$.

Similarly we can find flat directions (\ref{Dflat}) for the other $Z_{2n}$ 
orbifold models.
For example, the $U_3$ and $T_3$ sectors of $Z_6$-I orbifold models have 
the following superpotential; 
\begin{eqnarray}
W& =& \sum_a U_{3,P_U}T_{n,\tilde \Delta=1,a,P_1+nV}
T_{n,\tilde \Delta=1,a,P_2+nV} 
\nonumber \\
& + & \sum_b U_{3,P_U}T_{n,\tilde \Delta=\omega,b,-(P_1+nV)}
T_{n,\tilde \Delta=\omega^2,b,-(P_2+nV)},
\label{poten6}
\end{eqnarray}
where $a$ ($b$) denotes 6 (5) states with $e^{i\gamma}=1$ 
($\omega$ and $\omega^2$).
This potential also has the following flat direction;
\begin{eqnarray}
\langle T_{3,\tilde \Delta=1,a=c,P_c+nV}\rangle = 
\langle T_{3,\tilde \Delta=\omega,b=c,-(P_c+nV)}\rangle \neq 0,
\nonumber \\
\langle T_{3,\tilde \Delta=\omega^2,b,P_b+nV}\rangle = 
\langle T_{3,\tilde \Delta=\omega^2,b,-(P_b+nV)}\rangle \neq 0,
\label{flat6}
\end{eqnarray}
where $c=1 \sim 5$.
Along this direction a rank of a gauge group could break by at least 5.
In a similar way, other $T_n$ sectors of $Z_{2n}$ orbifold models 
can have flat directions.
Along these directions, ranks of gauge groups could reduce by the number of 
common degeneracy factors for conjugate pairs.

In addition to the flat directions for $T_n$, the $\hat U_3$ sector 
can have another flat direction as 
\begin{eqnarray}
\langle \hat U_{3,P} \rangle =\langle \hat U_{3,-P} \rangle 
\neq 0.
\label{flatu}
\end{eqnarray}
Because $W$ does not include $\chi (\hat U_3)^\ell$ couplings, 
even if we take into account nonrenormalizable couplings.
Thus this is a flat direction for all orders of $W$.
Therefore if a gauge group has these pairs of massless 
matter fields in the $\hat U_3$ sector, this group breaks reducing 
its rank by the number of pairs.
However both VEVs of $T_n$ and $\hat U_3$ (\ref{flat}) and 
(\ref{flatu}) are not always flat directions at the same time.
Because that could lead to 
$\langle \partial W/\partial T_n \rangle \neq 0$ for 
$W=\hat U_3(T_n)^2$.
Such a situation is model-dependent and 
we have to examine explicit $\hat U_3$ massless 
spectra of models.

Next we study effects of nonrenormalizable couplings on flat directions 
for $T_n$.
We discuss $\chi (T_n)^\ell$ couplings.
Among this type of couplings, the selection rule due to the point group 
allows $U_i(T_n)^{2\ell}$ couplings.
For example, $U_3(T_n)^{2\ell}$ couplings has the total $SO(6)$ momentum 
$(\ell-1,\ell-1,0)$ in $V_{-1/2}V_{-1/2}(V_{-1})^{2\ell-1}$ form.
We use the picture changing operator $(2\ell)$ times to change this 
form into $V_{-1/2}V_{-1/2}V_{-1}V_0^{2\ell-2}$ form, where 
the total $SO(6)$ momentum should be conserved for allowed couplings.
Thus the obtained couplings include the oscillators as 
\begin{eqnarray}
(\partial X_{1(n)})^{\ell -1}(\partial X_{2(n)})^{\ell -1}.
\end{eqnarray}
This should be invariant under the $Z_N$ twist of oscillators 
(\ref{osci}).
That requires $\ell-1=$even.
Thus the allowed couplings are obtained as 
$U_3(T_n)^{4m+2}$ couplings with $m=$ integer.
The other $U_i(T_n)^\ell$ couplings ($i=1,2$) are forbidden.
Even if we take into account these $U_3(T_n)^{4m+2}$ couplings, 
we find the flat direction where only one pair develop their 
VEVs.
Because in this case $U_3(T_{n,P+nV})^k(T_{n,-(P+nV)})^\ell$ couplings 
with $k+\ell=4m+2$ are dangerous couplings to lift the flatness, 
but we cannot obtain the nonzero $E_8$ root vector $P_U$ satisfying 
$P_U+k(P+nV)-\ell(P+nV)$ for the momentum $P+nV$ 
(\ref{T11}), (\ref{T2}) or their ``conjugates''.

Similarly we investigate flat direction in the case with more than one 
pairs developing VEVs.
For example, we discuss $U_3(T_2)^6$ couplings in $Z_4$ 
orbifold models.
Suppose that in Eq.(\ref{flat}) VEVs are developed by two pairs 
of states with 
\begin{eqnarray}
P_1+2V=(1,0,\cdots,0)(1/2,-1/2,0,\cdots,0),
\label{P1}
\end{eqnarray}
and $-(P_1+2V)$ for $c=1$, 
\begin{eqnarray}
P_2+2V=(0,1,0,\cdots,0)(1/2,-1/2,0,\cdots,0),
\label{P2}
\end{eqnarray}
and $-(P_2+2V)$ for $c=2$,
We assign these (\ref{P1}) and (\ref{P2}) to different 
6-dimensional ground states.
For simplicity, here we assign a conjugate pair of $(P_c+2V)$ and 
$-(P_c+2V)$ to the 
6-dimensional ground states with the same fixed points and 
the different values of $e^{i\gamma}$, e.g. 
$|e_2+ke_3 \rangle \pm |e_1+e_2+ke_3 \rangle$.
Then we investigate $U_3(T_2)^6$ couplings.
If these couplings include even fields of one conjugate pair, we 
can find no $U_3$ field to couple those fields.
Because $2m(P_c+2V)$ of Eqs. (\ref{T11}), (\ref{T2}) or 
their ``conjugates'' is too large compared with the $E_8$ root 
vectors (\ref{vector}) and (\ref{spinor}), which the $U_3$ states have, 
and by adding the other momenta, $-(P_1+2V)$, $(P_2+2V)$ and 
$-(P_2+2V)$, one cannot reduce it into the $E_8$ roots except 
vanishing momentum.
Note that the $U_3$ sector does not have vanishing momentum.
Thus nonvanishing $U_3(T_2)^6$ couplings should include odd fields of 
one conjugate pair.
For example, $(T_{2,c=1,P_1+2V})^2T_{2,c=1,-(P_1+2V)}
T_{2,c=2,P_2+2V}(T_{2,c=2,-(P_1+2V)})^2$ couplings have 
the $E_8 \times E'_8$ momentum $(1,-1,0,\cdots,0)(0.\cdots,0)$.
This momentum is included in the $E_8$ root vectors and the 
$U_3$ sector could have it in a certain case, although 
it is model-dependent whether a model really has this momentum.
We have to discuss the space group selection rules for this coupling.
Here corresponding fixed points are denoted as (\ref{fp})
\begin{eqnarray}
(\theta^2,i_1e_1+j_1e_2+k_1e_3+\ell_1 e_4) 
\quad {\rm for } \quad \pm (P_1+2V),\nonumber \\
(\theta^2,i_2e_1+j_2e_2+k_2e_3+\ell_2 e_4) 
\quad {\rm for } \quad \pm (P_2+2V).
\label{fp4}
\end{eqnarray}
Due to the space group selection rules (\ref{space}), these couplings 
are allowed if they satisfy 
\begin{eqnarray}
3(i_1+i_2)={\rm even}, \quad 3(j_1+j_2)={\rm even}, \\
\quad 3(k_1+k_2)={\rm even}, \quad 3(\ell_1+\ell_2)={\rm even}.
\end{eqnarray}
That implies $i_1=i_2$, $j_1=j_2$, $k_1=k_2$ and $\ell_1=\ell_2$, 
i.e., the exactly same fixed point.
Thus the different fixed points $c=1,2$ have no $U_3(T_2)^6$ 
coupling.
Similarly we can show that, in general, $T_n$ sectors of $Z_{2n}$ 
orbifold models have no $U_3(T_n)^{4m+2}$ couplings which 
include two conjugate pairs with any momenta of (\ref{T11}), 
(\ref{T2}) and their ``conjugates''.
Because for fixed points like Eq.(\ref{fp4}) this type of 
couplings requires 
\begin{eqnarray}
({\rm odd})\times i_1 +({\rm odd})\times i_2={\rm even},
\end{eqnarray}
and similar equations for $j$, $k$, and $\ell$, 
i.e., the same fixed point.
Thus for all orders the direction of Eqs. (\ref{Dflat}) and (\ref{flat}) 
is a flat direction, which reduces a rank of a gauge group by two.

We extend the above analysis to the case where three conjugate 
pairs develop their VEVs as Eq.(\ref{flat}) ($c=1,2,3$) 
in $Z_4$ orbifold models.
In addition to the fixed points (\ref{fp4}), we denote the third fixed 
point as 
\begin{eqnarray}
(\theta^2,i_3e_1+j_3e_2+k_3e_3+\ell_3 e_4), 
\end{eqnarray}
and its momenta as $\pm(P_3+2V)$.
In a similar way to the above discussion, $U_3(T_2)^{4m+2}$ 
couplings should include odd fields of each conjugate pair 
in $T_2$, due to the $E_8 \times E'_8$ momentum conservation.
Then the total number of the fields is odd.
That conflicts with $4m+2$.
Therefore such couplings are not allowed.
This can be extended into any case where odd conjugate pairs 
develop their VEVs in $T_n$ sectors of $Z_{2n}$ orbifold models.
In this case, $U_3(T_n)^{4m+2}$ couplings are not allowed.

Further we consider the case where four conjugate pairs develop 
their VEVs.
In addition to the above, we denote the fourth fixed point as 
\begin{eqnarray}
(\theta^2,i_4e_1+j_4e_2+k_4e_3+\ell_4 e_4), 
\end{eqnarray}
and its momentum $\pm(P_4+nV)$.
In this case, the space group selection rule as well as 
the $E_8 \times E'_8$ momentum conservation requires 
\begin{eqnarray}
\sum_{d=1}^4 ({\rm odd})\times i_d={\rm even},
\end{eqnarray}
and similar equations for $j_d$, $k_d$ and $\ell_d$.
We can find combinations of fixed points not to satisfy these 
equations, e.g.,
\begin{eqnarray}
(i,j,k,\ell)=(1,0,1,1),\  (1,0,1,0),\  (0,0,0,0) {\rm \ and \ } 
(0,0,1,1).
\end{eqnarray}
Such a combination leads to a flat direction, which break a gauge 
group reducing its rank by four.

Moreover we can find some combinations of fixed points not 
to satisfy the following equation;
\begin{eqnarray}
\sum_{d=1}^6 ({\rm odd})\times i_d={\rm even},
\end{eqnarray}
where $i_5$ and $i_6$ correspond to the fifth and sixth fixed points, 
or similar equations for $j_d$, $k_d$ and $\ell_d$.
Therefore the direction (\ref{flat}) with $c=1 \sim 6$ is 
still a flat direction for all orders of nonrenormalizable 
couplings $U_3(T_2)^{4m+2}$.
As results, gauge groups 
$\tilde G(E_7) \times \tilde G(SU(2)) \times \tilde G(E'_8)$ and 
$\tilde G(SO(16)) \times \tilde G(SU(2)') \times \tilde G(E'_7)$ 
in $Z_4$ orbifold models can break through the flat directions 
at least into $SU(2)_{E_7} \times \tilde G(E'_8)$ and 
$SU(3)_{SO(16)} \times  \tilde G(E'_7)$, where 
$SU(2)_{E_7}$ and $SU(3)_{SO(16)}$ denotes $SU(2)$ and $SU(3)$ 
subgroups in the original groups $E_7$ and $SO(16)$, respectively.
Similarly these gauge groups in other orbifold models break 
reducing their ranks by the numbers shown in the third column of 
Table 1 as the common number of degeneracy factors for 
conjugate pairs of $\tilde \Delta_n=\omega_{(n)}^k$ and 
$\tilde \Delta_n=(\omega_{(n)})^{-k-\delta}$.
Thus drastic symmetry breaking could happen.

These results have important phenomenological implications.
For example, we cannot expect appearance of the standard model
gauge group $SU(3)\times SU(2) \times U(1)$ in the $E_7$ sector 
of the $Z_4$ orbifold models with 
$\tilde G(E_7) \times \tilde G(SU(2)) \times \tilde G(E'_8)$.
In this case, we have to concentrate the $E'_8$ sector 
for candidates of the standard model gauge group.
On the other hand, ranks of gauge groups in $Z_8$ and 
$Z_{12}$ orbifold models are reduced not so much along 
the generic flat directions we have discussed because the degeneracy
factors are small compared with those of $Z_4$ and $Z_6$ orbifold 
models as given in Table 1.
For example, in these orbifold models it might be possible to 
derive the standard model gauge group in the $E_7$ sector from the 
$\tilde G(E_7) \times \tilde G(SU(2)) \times \tilde G(E'_8)$
model, although explicit models could have more accidental flat 
directions leading to further symmetry breaking.

\subsection{Example}

We investigate the previous results on the flat directions 
using an explicit model.
Here we take the $E_6 \times SU(2) \times U(1) \times \tilde G(E'_8)$ 
models (\ref{model1}) of $Z_4$ orbifold models.
As shown in Tables 2 and 4, this type of models have the following 
$T_2$ massless matter fields;
\begin{eqnarray}
10\times [(\overline {27},1)_1 +(1,1)_{-3}]
~~~(\tilde \Delta_2 =1,N^{(2)}_L=0),\\
6\times [(27,1)_{-1} + (1,1)_{3}]~~~(\tilde \Delta_2 =-1,N^{(2)}_L=0),\\
20\times (1,2)_0 ~~~(\tilde \Delta_2 =1,N^{(2)}_L=1/2),\\
12\times (1,2)_0 ~~~(\tilde \Delta_2 =-1,N^{(2)}_L=1/2),
\end{eqnarray}
and  the following $U_i$ massless matter fields;
\begin{eqnarray}
& U_1:(\overline {27},2)_1+(1,2)_{-3}, \quad 
U_2:(\overline {27},2)_1+(1,2)_{-3},\\
& U_3:(\overline {27},1)_{-2}+(27,1)_2,
\end{eqnarray}
under $ E_6 \times SU(2) \times U(1) $.
Further $U_i$'s include some $\tilde G(E'_8)$ matter fields 
without quantum numbers of $ E_6 \times SU(2) \times U(1) $.

The superpotential in $T_2$ and $U_3$ sectors is written as 
\begin{eqnarray}
W_{T_2T_2U_3} & = & \sum_a d_{ijk}(\overline {27}_{-2})_{U_3}^i
(\overline {27}_{1})_{T_2,a}^j(\overline {27}_{1})_{T_2,a}^k
+\sum_b d_{ijk}(27_{2})_{U_3}^i
(27_{-1})_{T_2,b}^j(27_{-1})_{T_2,b}^k \nonumber \\
 & + & \sum_a  (27_{2})_{U_3}^i(\overline {27}_{1})_{T_2,a}^i
(1_{-3})_{T_2,a}
+\sum_b  (\overline {27}_{-2})_{U_3}^i(27_{-1})_{T_2,b}^i
(1_{3})_{T_2,b},
\end{eqnarray}
where indices for $SU(2)$ are omitted, $(27)^i$ denotes the $i$-th 
element of a $27$ multiplet and $a$ and $b$ 
denote the $T_2$ states with $\tilde \Delta_2=1$ and $-1$, 
respectively.
Here $d_{ijk}$ denotes the third rank antisymmetric invariant.
This superpotential does not include the doublet fields.
Hence the doublet fields always develop their VEVs 
in the flat direction.

Let us study the following $D$-flat direction;
\begin{eqnarray}
<(\overline {27}_{-2})_{U_3}^i>=<(27_{2})_{U_3}^i>=v^i_0,\\
<(\overline {27}_{1})_{T_2,c(\Delta =1)}^i>=
<(27_{-1})_{T_2,c(\Delta =-1)}^i>=v^i_c,\\
<(1_{-3})_{T_2,c(\Delta =1)}>=
<(1_{3})_{T_2,c(\Delta =-1)}>=u_c,
\end{eqnarray}
where $c=1 \sim 6$.
We have the following conditions for the $F$-flatness;
\begin{eqnarray}
& <{ \partial W \over \partial (1)_{T_2,c}}>=v^i_0v^i_c=0,\\
& <{ \partial W \over \partial (27)^j_{T_2,c}}>=
d_{ijk}v^i_0v^k_c + v^j_0 u_c=0,\\
& <{ \partial W \over \partial (27)^i_{U_3}}>=
\sum_c [d_{ijk} v^j_cv^k_c +v^i_c u_c]=0.
\end{eqnarray}
Because of $d_{ijj}=0$, one of the flat directions is obtained as 
\begin{eqnarray}
v^i_0=0, \quad u_c=\delta^1_c, \quad v^i_c=\delta^i_c,
\end{eqnarray}
where $c \neq 1$ for the last equation.
Along this direction, the gauge group 
$E_6 \times U(1)$ can break into $SU(2)_{E6}$, reducing its rank 
by six.
That is consistent with the previous result.
On the top of that there is the flat direction where 
a combination of three $\overline {27}_{T_2,a}$ 
($a \neq c$ e.g. $a=7,8,9$ ) develops their VEVs at the same time.
In this case, $SU(2)_{E6}$ is also broken.
After these breaking, some of $T_1$ matter fields gain masses 
through the $T_1T_1T_2$ couplings 
and the observable $U_3$ matter fields gain masses.
However the $U_{1,2}$ matter fields remain massless.
There is another flat direction, where $v^i_0 \neq 0$.
In this case, all of the observable $U_{1,2}$ fields 
gain masses through the $U_1U_2U_3$ couplings.

We investigate more explicitly which fields become massive through 
the above flat directions, using the $Z_4$ orbifold model of No.1 
in Table 4 with the gauge group 
$E_6 \times SU(2) \times U(1) \times E'_8$.
As shown in Table 4, this model has the following $T_1$ massless fields;
\begin{eqnarray}
& 16 \times (\overline {27},1)_{-1/2}~~~(N^{(1)}_L=0),\\
& 32 \times (1,2)_{-3/2}~~~(N^{(1)}_L=1/4),\\
& 80 \times (1,1)_{3/2}~~~(N^{(1)}_L=1/2).
\end{eqnarray}

(1) $U(1)$ breaking

This happens when a singlet pair in $T_2$ develops VEVs, i.e. 
$u_c \neq 0$.
The singlet fields in $T_1$ gain masses through this breaking.
Among the 80 singlets, the 16 singlets remain massless and 
these corresponding $N^{(1)}_{Li}=(1/4,1/4,0)$.
Further through $W_{T_2T_2U_3}$ the two pairs of 
$(27 + \overline {27})$ of $U_3$ and $T_2$ 
obtain mass terms.

(2)$E_6 \rightarrow SO(10)$ breaking

Case I: This happens when a $(27 + \overline {27})$ pair 
in $T_2$ develop VEVs, i.e. $v^1_c \neq 0$.
We assign the first element $(27)^1$ 
as the $SO(10)$ singlet $1_4$.
Then $1_4$ in $27$ and $1_{-4}$ in $\overline {27}$ 
develop VEVs.
Because of $d_{i11}=0$, this satisfies the flatness 
conditions.
Through this breaking, every $27$ and $\overline {27}$ 
are decomposed into
\begin{eqnarray}
27=1_4+10_{-2} +16_1, \quad 
\overline {27}=1_{-4} +10_2 + {\overline{16}}_{-1},
\end{eqnarray}
under $SO(10) \times U(1)_{E6}$.
Through this breaking, every generation of the 10-fields in $T_1$ 
has a mass by $(T_1)^2T_2$ couplings.
There appear fourteen massless singlets without any 
quantum numbers for unbroken gauge groups in $T_2$.
Note that for the whole gauge group nonvanishing $v^1_c$ 
leads to $E_6 \times U(1) \rightarrow SO(10) \times U(1)_I$,
where the current of $U(1)_I$ is a linear combination of 
those for $U(1)_{E6}$ and the original $U(1)$, i.e., 
$J_{U(1)_I}=J_{U(1)E6}+4J_{U(1)}$.

One can break $U(1)_I$ with nonvanishing $u_c$ as (1). 
In this case the flatness conditions require 
$\sum_c v^1_cu_c=0$.
That implies the Higgs fields 
breaking $E_6$ should sit on the different fixed point 
from one for the Higgs fields breaking $U(1)$.
At this stage we have the following massless $U_i$ spectrum 
under the unbroken $SO(10)\times SU(2)$;
\begin{eqnarray}
U_1:(\overline {27},2)+(1,2), \quad U_2:(\overline {27},2)+(1,2),
\end{eqnarray}
the following massless $T_1$ spectrum;
\begin{eqnarray}
& 16\times [(1,1) + (\overline {16},1)](N^{(1)}_L=0), \\
& 32\times (1,2) (N^{(1)}_L=1/4), \quad 
16\times (1,1) (N^{(1)}_L=1/2),
\end{eqnarray}
and the following massless $T_2$ spectrum;
\begin{eqnarray}
& 8\times (\overline {27},1) (\Delta =1,N^{(2)}_L=0), \quad 
4\times (27,1) (\Delta =-1,N^{(2)}_L=0), \\
& 9\times (1,1) (\Delta =1,N^{(2)}_L=0), \quad 
5\times (1,1) (\Delta =-1,N^{(2)}_L=0), \\
& 20\times (1,2) (\Delta =1,,N^{(2)}_L=1/2), \quad 
12\times (1,2) (\Delta =-1,N^{(2)}_L=1/2), 
\end{eqnarray}
where $27$ is a short notation for 
$1+10 +16$.

Case II:This happens when a $(27 + \overline {27})$ pair 
in $U_3$ develops VEVs, i.e. $v^1_0 \neq 0$.
In a similar way to Case I, we can calculate massive modes.
In this case, every 10-field in $U_i$ and $T_2$ becomes 
massive.
In addition, $(1,2)_{-3}$ and the $SO(10)$ singlets in $\overline {27}$ 
of $U_{1,2}$ become massive.
The nonvanishing $v^1_0$ leads to 
the breaking $E_6 \times U(1)\rightarrow SO(10)\times U(1)_J$, 
where the current $J_{U(1)_J}$ is obtained as 
$J_{U(1)_J}= J_{U(1)E6}-2J_{U(1)}$.
Let us study the $U(1)_J$ breaking by nonvanishing $u_c$.
In this case, however, the flatness conditions 
can not be satisfied with nonvanishing 
$v^j_0$ and $u_c$ at the same time.
Thus in Case II there is no $U(1)_J$ breaking.

(3) $SO(10) \rightarrow SU(5)$ breaking

For the $SO(10)$ model without $U(1)_I$ obtained in Case I, we 
consider further breaking, where 
a pair of $(16 + \overline {16})$ develops VEVs, i.e. 
$v^1_{c=1}\neq 0$, $v^2_{c=2} \neq 0$ and $u_{c=3} \neq 0$ 
at the same time, where $v^2_2$ corresponds to 
a $SU(5)$ singlet in 16.
Because of $d_{i12}=d_{i22}=0$, these VEVs satisfy the 
flatness conditions.

Through this breaking, the gauge group breaks as 
$SO(10) \rightarrow SU(5)$.
The representations $16$ and $\overline {16}$ are decomposed as 
\begin{eqnarray}
16=1_{-5}+\overline 5_3 + 10_{-1}, \quad 
\overline {16}=1_5 + 5_{-3} + \overline {10}_1,
\end{eqnarray}
under $SU(5) \times U(1)_{SO(10)}$.
Further 10 representation is decomposed as 
\begin{eqnarray}
10=5_2 + \overline 5_{-2}.
\end{eqnarray}
Through this breaking, no matter fields obtain masses except the 
Higgs fields.
For example, the $\overline {16}$-fields in $T_1$ 
cannot couple through $T_1T_1T_2$-coupling 
with the 16 or $\overline {16}$- field in $T_2$, to whose  
multiplet the Higgs field belongs.
Thus the massless spectrum is changed not so drastically.
In the above notation, 
27 means under this gauge group as 
\begin{eqnarray}
27=1+1+5+ \overline 5 +\overline 5 +10.
\end{eqnarray}

We can continue this symmetry breaking as 
$$SU(5) \rightarrow  SU(4) \rightarrow SU(3) \rightarrow SU(2),$$
using a 5 representation of subgroup $SU(5)$ in 16-multiplet of $SO(10)$, 
4 and 3 representations of subgroups $SU(4)$ and $SU(3)$, respectively.
Also this $SU(2)$ can be broken by one pair of doublets.
Through this series of breakings,  the number of $T_1$ 
massless field is not changed, since $16(\overline {16})^2$ 
coupling is forbidden.
For $T_2$, the number of $27$ representation is reduced by one 
leaving singlets under unbroken groups, as the rank of 
the gauge group is reduced by one.

Alternatively we can reduce the number of the $T_1$ massless 
fields in different types of breakings.
For example, we can break $SU(5)$ by 10($=5+{\overline 5}$ 
in $SU(5)$ base) of the $SO(10)$ multiplet in $T_2$.
This allows the $10(\overline {16})^2$ coupling.
In this breaking, we have the gauge group 
$SU(3) \times SU(2)$ and some of $T_1$ fields 
gain masses.
We can investigate flat directions of other models explicitly.

\subsection{Wilson lines}

In addition to $V^I$, Wilson lines $a^I_{e^i}$ can be embedded 
into $\Gamma_{E_8 \times E'_8}$ as \cite{WL1,WL2,Orbi4}
\begin{eqnarray}
(\theta^k,e^i) \rightarrow (kV^I,a^I_{e^i}).
\label{WL}
\end{eqnarray}
The lattice vectors related through $\theta$ should 
correspond to an equivalent Wilson line, i.e. 
$a^I_{e^i}=a^I_{(\theta e)^i}$.
For example, the first equation of (\ref{Cox1}) leads to the 
constraint that $2a^I_{e_a}$ ($a=2,4$) should be on 
$\Gamma_{E_8 \times E'_8}$ in $Z_4$ orbifold models.
Further the second equation requires $a^I_{e_a}=0$ 
(mod $\Gamma_{E_8 \times E'_8}$) ($a=1,3$).
Similarly Eq.~(\ref{Cox2}) leads to the conditions 
$2a^I_{e_5}=2a^I_{e_6}=0$ (mod $\Gamma_{E_8 \times E'_8}$).
It is remarkable that the states in the same linear 
combination to construct a $\theta$-eigenstate 
like Eq.~(\ref{state}) have the same Wilson lines.
The structure of Wilson lines depends on the Lie lattice to 
construct the orbifold, because $\theta$ transforms 
lattice vectors in a different way.
Appendix A shows the Lie lattice leading to Wilson lines 
with the most degrees of freedom among the same $Z_N$ orbifold 
models \cite{WL2,Orbi4}.

For the models with Wilson lines $a^I$, the massless $U$ fields 
should satisfy the following equation;
\begin{eqnarray}
\sum_IP^Ia^I={\rm integer}.
\label{WL1}
\end{eqnarray}
These Wilson lines could lead to smaller gauge groups and 
reduce the number of massless $U$ matter fields.
Hence Wilson lines are important to derive realistic models.
If $P^I$ satisfies Eq.(\ref{WL1}), its conjugate $-P^I$ also 
satisfies it.
Thus the $\hat U_3$ sector can have conjugate pairs, $R$ and 
$\overline R$.

The $T_k$ sector corresponding to Eq.(\ref{WL}) has the shifted 
momentum $(P+kV+a_{e^i})$.
Its left-moving massless condition is written as 
\begin{eqnarray}
{1 \over 2} \sum_{I=1}^{16} (P^I+kV^I+a^I_{e^i})^2+N^{(k)}_L
+c_k-1=0.
\label{WLmass}
\end{eqnarray}
Thus Wilson lines resolve degeneracy of $T_k$ massless matter fields.
The GSO phase with Wilson lines is obtained by replacing 
$kV^I$ in Eq.(\ref{GSO'}) into $kV^I+a^I_{e^i}$.
Further the constraint due to the modular invariance 
is written as
\begin{eqnarray}
N\sum_{i=1}^3(kv^i)^2-N\sum_{I=1}^{16}(kV^I+a^I_{e^i})^2
={\rm even},
\label{WLmod}
\end{eqnarray}
for each value of $k$.

Now let us consider conjugate pairs $R$ and $\overline R$ in 
the $T_2$ sector of $Z_4$ orbifold models.
The fixed points of Eq.(\ref{fp}) have the following 
$E_8 \times E'_8$ momenta;
\begin{eqnarray}
P^I+2V^I+ja^I_{e_2}+\ell a^I_{e_4}.
\end{eqnarray}
Since both of $a^I_{e_2}$ and $a^I_{e_4}$ are Wilson lines 
with the order 2, we have 
\begin{eqnarray}
\tilde V^I_{j,\ell} \equiv 2V^I+ja^I_{e_2}+\ell a^I_{e_4}, \quad 
2\tilde V^I_{j,\ell}=0 {\rm \ (mod \ } \Gamma_{E_8 \times E'_8}).
\end{eqnarray}
That implies if $P^I+\tilde V^I_{j,\ell}$ satisfies the massless 
condition (\ref{WLmass}), we always have 
$P'=-P^I-\tilde V^I_{j,\ell}$ which sits on $\Gamma_{E_8 \times E'_8}$, 
and $P'^I+\tilde V^I_{j,\ell}$ satisfies the same massless condition 
for each combination of $(j,\ell)$.
The transformation of $\tilde \Delta_n$ under charge conjugation 
is obtained in a way similar to the case without Wilson lines.
Thus the $T_2$ sector can have conjugate pairs $R$ and $\overline R$ 
for the states with same values of $(j,\ell)$.
It is notable that these conjugate pairs appear in the  states of 
linear combinations with the same content and different 
eigenvalues, e.g., $|e_2+ke_3\rangle +|e_1+e_2+ke_3\rangle$ and 
$|e_2+ke_3\rangle -|e_1+e_2+ke_3\rangle$, while in the case 
without Wilson lines we can make pairs of any combinations with 
$\tilde \Delta_2=1$ and $-1$ due to the degeneracy.
Further generally conjugate pairs do not appear for the states with 
$\tilde \Delta_2=1$ 
which have no associated states with $\tilde \Delta_2=-1$, e.g.  
$|ie_1 +ke_3 \rangle$.

Since $\tilde V^I_{j,\ell}$ is a shift with the order 2, each of 
its $E_8$ part ($I=1 \sim 8$) and its $E'_8$ part ($I=9 \sim 16$) 
is classified three classes, Classes 1, 2 and 3 in the same way 
as the case without Wilson lines.
Further the modular invariance (\ref{WLmod}) allows only two 
combinations of these Classes for $\tilde V^I_{j,\ell}$, 
Classes 2 \& 1 and Classes 3 \& 2.
Thus the massless $T_2$ fields have the shifted $E_8 \times E'_8$ 
momenta of Eqs. (\ref{T11}), (\ref{T12}), (\ref{T2}) 
and their ``conjugates'' for each $(j,\ell)$.
Note that in one model the twisted matter fields of 
Classes 2 \& 1 and Classes 3 \& 2 appear generally for different 
values of $(j,\ell)$.
However every $Z_4$ orbifold models with Wilson lines have 
conjugate pairs, $R$ and $\overline R$, in the $T_2$ sector.
These pairs lead to a flat direction in a way similar to the case 
without Wilson lines.
This flat direction could break a gauge group reducing its rank by 6.

Here we discuss examples of models with Wilson lines.
We take the $Z_4$ orbifold model with the following shift $V^I$;
\begin{eqnarray}
V^I={1 \over 4}(2,1,1,0,\cdots,0)(0,\cdots,0).
\label{WLshift}
\end{eqnarray}
In one example, we consider the above $Z_4$ orbifold model with 
the following Wilson line;
\begin{eqnarray}
a^I_{e_2}=(0,\cdots,0)(1,0,\cdots,0).
\end{eqnarray}
This Wilson line is associated with the lattice vector $e_2$.
This combination of $V^I$ and $a^I_{e_2}$ is consistent with 
the modular invariance (\ref{WLmod}).
We obtain $E_6 \times SU(2) \times U(1) \times SO(16)'$ gauge group 
by $V^I$ and $a^I_{e_2}$.
This orbifold model has the following massless $\hat U_3$ fields;
\begin{eqnarray}
(\overline {27},1;1) + (27,1;1) + (1,1;128_s).
\end{eqnarray}
The other untwisted subsectors, $U_1$ and $U_2$, are found as the 
fourth column for the \# 3 shift in Table 2.
Among the $T_2$ states, the following states;
\begin{eqnarray}
|ie_1+e_4 \rangle \pm |ie_1+e_3+e_4 \rangle, \quad (i=0,1)
\label{state41}
\end{eqnarray}
as well as $|ie_1 +ke_3 \rangle$ ($i,k=0,1$) have no Wilson line and 
their momenta are obtained as $P^I+2V^I$.
This shift corresponds to Classes 2 \& 1.
Thus the massless states of Eq.(\ref{state41}) have Eqs.(\ref{T11}), 
(\ref{T12}) and their ``conjugates'' as the shifted 
$E_8 \times E'_8$ momenta, 
leading to two conjugate pairs, $R$ and $\overline R$.
The other $T_2$ states;
\begin{eqnarray}
& |e_2 +ke_3\rangle \pm |e_1+e_2 +ke_3\rangle, \quad 
|e_2+e_4 \rangle \pm |e_1+e_2+e_3+e_4 \rangle, \nonumber \\
& |e_2+e_3+e_4 \rangle \pm |e_1+e_2+e_4 \rangle,
\label{state42}
\end{eqnarray}
have the momenta $P^I+2V^I+a^I_{e_2}$.
Note that $2V^I+a^I_{e_2}$ corresponds to Classes 2 \& 3.
These massless states (\ref{state42}) have Eq.(\ref{T2}) and 
their ``conjugates'' as the shifted $E_8 \times E'_8$ momenta, 
leading to four conjugate pairs, $R$ and $\overline R$.
Note that these four conjugate pairs have quantum numbers under 
$SO(16)'$.
Therefore the $E_6 \times U(1)$ group is broken by VEVs of 
two conjugate pairs associated with Eq.(\ref{state41}) and 
its rank reduces by 2.
The $SU(2)$ group also breaks.
VEVs of four conjugate pairs corresponding to Eq.(\ref{state42}) 
reduce the rank of the $SO(16)'$ group by four.
Further VEVs of the 27 and $\overline {27}$ ($128'_s$) in 
$\hat U_3$ reduce the rank of the $E_6 \times U(1)$ ($SO(16)'$) 
group by one.

We discuss another example.
We consider the $Z_4$ orbifold model with above shift 
(\ref{WLshift}) and the following Wilson line;
\begin{eqnarray}
a^I_{e_2}=(0,1/2,1/2,0,\cdots,0)(1/2,1/2,0,\cdots,0).
\end{eqnarray}
This combination of $V^I$ and $a^I_{e_2}$ is consistent with 
the modular invariance (\ref{WLmod}).
This model has the 
$SO(10) \times SU(2) \times U(1)^2 \times E'_7 \times SU(2)'$ 
gauge group.
This model has the following $\hat U_3$;
\begin{eqnarray}
(27,1;1,1) + (\overline {27},1;1,1) + (1,1;56,2)
\end{eqnarray}
where we use short notations 27 and $\overline{27}$ in place of 
$27=1+10+16$ and $\overline{27}=1+10+\overline{16}$, respectively.
The states of Eqs.(\ref{state41}) have no Wilson lines and they have 
Eqs.(\ref{T11}), (\ref{T12}) and their ``conjugates'' 
as the $P^I+2V^I$.
The states of Eq.(\ref{state42}) have the momenta 
$P^I+2V^I+a^I_{e_2}$, where $2V^I+a^I_{e_2}$ corresponds to 
Classes 3 \& 2.
The shifted momenta $P^I+2V^I+a^I_{e_2}$ for massless states are 
written as Eq.(\ref{T2}) and their ``conjugates'', 
including four conjugate pairs.
Thus VEVs of these conjugate pairs reduce the rank of 
$SO(10) \times U(1)^2$ by 6.
Also $SU(2)$ and $SU(2)'$ are broken.
VEVs of the $\hat U_3$ sector can break  
$SO(10) \times U(1)^2$ and $E'_7$ reducing their ranks by one.

Similarly we can study $T_n$ sectors of the other $Z_{2n}$ orbifold 
models with nontrivial Wilson lines.
Especially $T_3$ of $Z_6$-I and $T_6$ of $Z_{12}$-I and II orbifold 
models cannot have nontrivial Wilson lines as shown in Appendix A, 
although other sectors of these orbifold models have nontrivial 
Wilson lines.
For these orbifold models, situation on flat directions 
is the same as the case without Wilson lines.
In general, only Wilson lines with the order 2 are allowed for 
$T_n$ sectors of the other $Z_{2n}$ orbifold models 
as shown in Appendix A.
Hence $nV^I+a^I_{e_i}$ is a shift vector with the order 2.
Its $E_8$ part is classifies into three classes, 
Classes 1, 2 and 3.
The modular invariance allows only the two combinations, 
Classes 2 \& 1 and 3 \& 2.
Thus massless matter fields of $T_n$ are obtained by 
Eqs.(\ref{T11}), (\ref{T12}), (\ref{T2}) and their ``conjugates''.
Conjugate pairs appear for the states by linear combination of the 
same content with values of $\tilde \Delta_n$ related by the 
conjugation in Table 1.
VEVs of these pairs could lead to flat directions in a way
similar to the case without Wilson lines.
Therefore a drastic symmetry breaking could also happen 
in the presence of nontrivial Wilson lines.

\subsection{Anomalies}

Along flat directions, several fields become massive 
as seen in subsection 4.2.
It is interesting to study flow of universal indices of models 
such as anomaly coefficients along these flat directions 
if any universal indices exist.
In superstring theories, anomalies as mixed $U(1)$ anomalies 
and mixed target-space duality anomalies can be canceled 
by the Green-Schwarz mechanism \cite{GS}.
This mechanism is independent of gauge groups.
Thus these mixed anomaly coefficients should be universal for gauge 
groups, although target-space duality anomaly coefficients are 
not always universal for unrotated planes under some twists 
because of another group-dependent cancellation mechanism, i.e., 
threshold corrections due to massive modes \cite{Thres,DFKZ}.
It is obvious that $U(1)$ anomaly coefficients do not change 
along flat directions if the anomalous $U(1)$ is not broken.
Suppose that the anomalous $U(1)$ is unbroken by VEVs.
Two fields in a mass term which is generated by flat directions 
have opposite anomalous $U(1)$ charges.
Then integrating out these fields does not contribute to the 
anomalous $U(1)$ coefficients.

Now let us discuss the target-space duality anomaly coefficients 
$b'^i_a$ for the $i$-th plane, which are obtained as 
\begin{eqnarray}
b'^i_a=-C(G_a)+\sum_{R_a}T(R_a)(1+2n^i_{R_a}),
\label{danom}
\end{eqnarray}
where $C(G_a)$ and $T(R_a)$ are Casimir for the adjoint 
representation and the index for the $R_a$ representation.
Here $n^i_{R_a}$ is a modular weight of the state with $R_a$ 
for the $i$-th moduli field \cite{weight}.
For completely rotated planes, 
they should be independent of gauge groups, i.e. 
$b'^i_a=$constant (for $a$) so that 
the Green-Schwarz mechanism works \cite{DFKZ}.
This leads to a strong constraint on massless spectra, which 
is phenomenologically interesting \cite{IL,anomcon}.
In Ref.\cite{IL}, it is shown that flat directions along the $U$ 
sectors do not change these duality anomaly coefficients 
if these $U$ sectors correspond to completely rotated planes.
Because these $U$ sectors consist in $N=4$ SUSY multiplets.
It is notable that the $T_n$ sectors of $Z_{2n}$ orbifold models 
do not contribute to the duality anomaly (\ref{danom}) for 
completely rotated planes, because these sectors have 
$n^i=(-1/2,-1/2,0)$ and the third plane is unrotated under 
$\theta^n$.
Hence integrating out these fields has no effect on 
the relation of duality anomaly coefficients.
However, if other $T_k$ sectors gain masses by VEVs of $T_n$, 
integration of these $T_k$ sectors changes Eq.(\ref{danom}).
In general, $b'^i_a$ are not universal for all unbroken gauge 
groups.

The flat directions are discussed in the presence of anomalous
$U(1)$ symmetry ($U(1)_A$) \cite{STflat,STflat2}.
Its $D$-term is given as \cite{V_DA,ST-FI},
\begin{eqnarray}
D^A &\equiv& {\delta_{GS}^{A} \over S+S^*} 
+ q_\kappa |\Phi^\kappa |^2, 
\label{V(DA)}
\end{eqnarray}
where $\delta_{GS}^{A}$ is a coefficient of the Green-Schwarz
mechanism to cancel the $U(1)_A$ anomaly \cite{GS} 
and $q_\kappa$ is a $U(1)_A$ charge of a 
matter multiplet $\Phi^{\kappa}$.
The dilaton field $S$ transforms nontrivially as 
$S \rightarrow S-i\delta_{GS}^{A}\theta(x)$ under $U(1)_A$ 
with the transformation parameter $\theta(x)$.
This nontrivial transformation generates the contribution 
of $S$ in the $D$-term (\ref{V(DA)}).
We have a relation $\langle S \rangle = 1/k_ag_a^2$
where $g_a$'s are the gauge coupling constants for 
gauge groups $G_a$ and $k_a$ are corresponding Kac-Moody levels.
The $D$-flatness of the anomalous $U(1)$ symmetry implies 
$\langle q_\kappa |\Phi^\kappa |^2 \rangle \neq 0$ 
for a finite value of $\langle S \rangle$, although 
the flat directions we have discussed 
lead to $\langle q_\kappa |\Phi^\kappa |^2 \rangle = 0$.
Thus we need another type of VEVs leading to 
$\langle q_\kappa |\Phi^\kappa |^2 \rangle \neq 0$, 
the other vanishing $D$-terms and $F$-flatness 
in order to obtain a finite value for $\langle S \rangle$.
Within the presence of such VEVs, the $D$-flatness of 
$\langle R \rangle = \langle {\bar R} \rangle$, i.e., 
Eq. (\ref{Dflat}), is obvious.
We can also investigate their $F$-flatness using explicit 
superpotentials.
In general, one model has several combinations of VEVs leading to 
$\langle q_\kappa |\Phi^\kappa |^2 \rangle \neq 0$, 
the other vanishing $D$-terms 
and $F$-flatness \cite{STflat,STflat2,STflat3}.
Some of such combinations might be inconsistent with the 
flat directions which we have discussed.

\section{Conclusions and discussions}

We have studied generic features related to matter contents and
flat directions in $Z_{2n}$ orbifold models.
We have found the existence of model-independent conjugate pairs in the 
$\hat U_3$ and $T_n$ sectors.
We have classified a number of $Z_{2n}$ orbifold models into 
only two types using the twisted sectors and the modular invariance.

Further applications of this classification are expected.
For example, we can classify $T_2$ ($T_4$) sectors of $Z_6$ 
($Z_{12}$) orbifold models in the similar way.
These massless fields have the same $E_8 \times E'_8$ momenta 
as $T_1$ of $Z_3$ orbifold models.
Further we can classify $T_2$ ($T_3$) sectors of $Z_8$ ($Z_{12}$) 
orbifold models.

Moreover these classification is very useful for 
$Z_{2n} \times Z_{2m}$ orbifold models with two independent 
twists, $\theta$ and $\omega$.
We can show that conjugate pairs, $R$ and $\overline R$, appear 
in massless spectra of three twisted sectors, i.e., 
$\theta^n$-twisted sector, $\omega^m$-twisted sector and 
$\theta^n\omega^m$-twisted sector.
These massless fields have the same shifted $E_8 \times E'_8$ 
momenta as the $T_n$ sectors of $Z_{2n}$ orbifold models.
It is interesting to study flat directions for these sectors of 
$Z_{2n} \times Z_{2m}$ orbifold models in a way similar to 
the above.

It has been also shown that this classification is very useful for
study on the breakings by flat directions.
We have shown that there exist generic flat directions 
in $T_n$ and $\hat{U}_3$ sectors. 
Conjugate pairs, $R$ and $\overline R$, lead to $D$-flatness as 
$\langle R \rangle = \langle \overline R \rangle \neq 0$.
These VEVs, in general, lead to flat directions and 
break gauge groups drastically.
These results are very important from phenomenological viewpoint.
There can be another type of flat directions in other sectors, but
we have not studied their features because the study requires a
model-dependent analysis and is beyond the aim of this paper.
The search for a realistic model based on $Z_{2n}$ orbifold models 
has been worth challenging and our results should be extended further.

Other string models, e.g., by Calabi-Yau construction or 
fermionic construction include some conjugate pairs, $R$ and $\overline R$.
It is interesting to apply the above analyses to these models.

Much work is recently devoted to derive soft SUSY 
breaking terms from superstring theory \cite{ST-soft,ST-soft2,ST-soft3}.
In these aspects, study on flat directions are also important.
For example, SUSY breaking in string models could lead to imaginary 
soft scalar masses at the Planck scale.
Such initial conditions at the Planck scale are obtained in the case 
where moduli fields contribute on the SUSY breaking 
in a sizable way \cite{ST-soft2}.
Imaginary soft scalar masses are also obtained in string models 
with anomalous $U(1)$ symmetries, even though the dilaton field 
breaks dominantly the SUSY \cite{ST-soft3}.
Such negative mass squared terms 
along flat directions have important implications.
Further SUSY breaking effects in supergravity might lift 
some flat directions and fix magnitudes of VEVs of matter 
multiplets determining gauge symmetry breaking pattern.
In the case that there are flat directions in the SUSY limit,
study of the structure of soft SUSY breaking terms and
its phenomenological implications
will be discussed elsewhere \cite{soft+flat}.

Other nonperturbative effects could lift these flat directions.
In Ref.\cite{BD}, it is shown nonperturbative effects like 
$e^{-aS}$ are forbidden in some Calabi-Yau models with 
an anomalous $U(1)$ symmetry due to holomorphy of $W$ and 
discrete symmetries.
It is important to apply such analyses to the generic 
flat directions obtained in this paper.

\section*{Appendix A}
Here we summarize the structure of $T_n$ sectors of $Z_{2n}$ 
orbifold models.
In (a), $v^i$ and the Lie lattice realizing it are shown.
This lattice has the most degrees of freedom for Wilson lines 
among the Lie lattices leading to the same value of $v^i$.
Transformation of vectors by $\theta$ is written in (b) \footnote{
For the $Z_4$ orbifold model, (b) is omitted. 
See Eqs.~(\ref{Cox1}) and (\ref{Cox2}).}.
In the following lattices, fixed points of $T_n$ sectors are 
$(\theta^2,ie_1+je_2+ke_3+\ell e_4)$  ($i,j,k,\ell=0,1$).
Eigenstates of $T_n$ are given in (c), where $\theta^n$ is 
omitted \footnote{Eigenstates written in Ref.\cite{Orbi4} 
include a minor mistake.
That should be replaced.}.
Constraints on Wilson lines are written in (d), where 
$a_b$ denotes $a^I_{e_b}$ and $Ma_b\approx 0$ means 
$Ma_b=0$ (mod $\Gamma_{E_8 \times E'_8}$).

\subsection*{A.1 $Z_4$ orbifold model}

\paragraph{(a)}
 $v^i=1/4(1,1,-2)$ 
\hskip 1cm
$SO(5)^2\times SU(2)^2$ lattice

\paragraph{(c)}Eigenstates
\begin{eqnarray*}
&|ie_1+ke_3\rangle\\
&|e_2+ke_3\rangle \pm |e_1+e_2+ke_3\rangle ,
\quad 
|ie_1+e_4\rangle \pm |ie_1+e_3+e_4\rangle,\\
&|e_2+e_4 \rangle \pm |e_1+e_2+e_3+e_4 \rangle,
\quad 
|e_2+e_3+e_4\rangle \pm |e_1+e_2+e_4\rangle.
\end{eqnarray*}

\paragraph{(d)}Wilson lines \ 
$2a_2\approx 2a_4 \approx 2a_5 \approx 2a_6 \approx 0,$ \ 
($a_1 \approx a_3 \approx 0$).

\subsection*{A.2 $Z_6$-I orbifold model}

\paragraph{(a)}
 $v^i=1/6(1,1,-2)$ 
\hskip 1cm
$G_2^2 \times SU(3)$ lattice

\paragraph{(b)} Twist
\begin{eqnarray*}
&\theta e_a=2e_a+3e_{a+1}, \quad
\theta e_{a+1}=-e_a-e_{a+1}, \quad 
a=1,3,\\
&\theta e_5=e_6, \quad 
\theta e_6=-e_5-e_6.
\end{eqnarray*}

\paragraph{(c)}Eigenstates
\begin{eqnarray*} 
|0\rangle, \quad 
|e_a\rangle + \alpha |e_a+e_{a+1}\rangle 
+ \alpha^2 |e_{a+1}\rangle, \quad (a=1,3),\\
|e_1+e_3\rangle + \alpha |e_1+e_2+e_3+e_4\rangle
+\alpha^2 |e_2+e_4\rangle,\\
|e_1+e_3+e_4\rangle + \alpha |e_1+e_2+e_4\rangle
+\alpha^2 |e_2+e_3\rangle,\\
|e_1+e_4\rangle + \alpha |e_1+e_2+e_3\rangle
+\alpha^2 |e_2+e_3+e_4\rangle,
\end{eqnarray*}
where $\alpha={\rm exp}[2\pi im/3]$ and $m=0,1,2$.

\paragraph{(d)}Wilson lines \ 
$3a_5\approx 0$, \ ($a_5 \approx a_6$), \ 
$a_b \approx 0$ \ ($b=1 \sim 4$).

\subsection*{A.3 $Z_6$-II orbifold model}

\paragraph{(a)}
 $v^i=1/6(1,-3,2)$ 
\hskip 1cm
$G_2\times SU(2)^2 \times SU(3)$ lattice

\paragraph{(b)} Twist
\begin{eqnarray*}
&\theta e_1=2e_1+3e_2, \quad
\theta e_2=-e_1-e_2, \quad 
\theta e_a=-e_a, \quad a=3,4 \\
&\theta e_5=e_6, \quad 
\theta e_6=-e_5-e_6.
\end{eqnarray*}

\paragraph{(c)}Eigenstates
\begin{eqnarray*} 
&|ke_3+\ell e_4 \rangle,\\
&|e_1+ke_3+\ell e_4 \rangle 
+ \alpha |e_1+e_2+ke_3+\ell e_4 \rangle
+ \alpha^2 |e_2+ke_3+\ell e_4 \rangle,
\end{eqnarray*}
where $\alpha={\rm exp}[2\pi im/3]$ and $m=0,1,2$.

\paragraph{(d)}Wilson lines \ 
$2a_3 \approx 2a_4 \approx 3a_5\approx 0$, \ 
($a_5 \approx a_6$, \ $a_1 \approx a_2 \approx 0$).

\subsection*{A.4 $Z_8$-I orbifold model}

\paragraph{(a)}
 $v^i=1/8(1,-3,2)$ 
\hskip 1cm
$SO(9) \times SO(5)$ lattice

\paragraph{(b)} Twist
\begin{eqnarray*}
&\theta e_1=e_2, \quad \theta e_2=e_3, \quad 
\theta e_3=e_1+e_2+e_3+2e_4, \\
&\theta e_4=-\sum_{a=1}^4e_a, \quad 
\theta e_5=e_5+2e_6, \quad 
\theta e_6=-e_5-e_6.
\end{eqnarray*}

\paragraph{(c)}Eigenstates
\begin{eqnarray*} 
&|i(e_1+e_3)\rangle, \quad 
|e_1+e_2\rangle \pm |e_2+e_3\rangle,\\
&|e_1\rangle + \alpha |e_1+e_2+e_3\rangle 
+\alpha^2 |e_3\rangle +\alpha^3 |e_2\rangle,\\
&|e_4\rangle + \alpha |e_3+e_4\rangle 
+\alpha^2 |e_2+e_3+e_4\rangle 
+\alpha^3 |\sum_{a=1}^4e_a\rangle,\\
&|e_1+e_4\rangle + \alpha |e_1+e_2+e_4\rangle 
+\alpha^2 |e_2+e_4\rangle 
+\alpha^3 |e_1+e_3+e_4\rangle,\\
\end{eqnarray*}
where $\alpha={\rm exp}[\pi im/2]$ and $m=0,1,2,3$.

\paragraph{(d)}Wilson lines \ 
$2a_4 \approx 2a_6\approx 0$, \ 
($a_1 \approx a_2 \approx a_3 \approx a_5 \approx 0$).

\subsection*{A.5 $Z_8$-II orbifold model}

\paragraph{(a)}
 $v^i=1/8(1,3,-4)$ 
\hskip 1cm
$SO(9) \times SU(2)^2$ lattice

\paragraph{(b)} Twist
\begin{eqnarray*}
&\theta e_1=e_2, \quad \theta e_2=e_3, \quad 
\theta e_3=e_1+e_2+e_3+2e_4, \\
&\theta e_4=-\sum_{a=1}^4e_a, \quad 
\theta e_b=-e_b, \quad b=5,6.
\end{eqnarray*}

\paragraph{(c)}Eigenstates are same those of $T_4$ for the $Z_8$-I 
orbifold models.

\paragraph{(d)}Wilson lines \ 
$2a_4 \approx 2a_5 \approx 2a_6\approx 0$, \ 
($a_1 \approx a_2 \approx a_3 \approx 0$).

\subsection*{A.6 $Z_{12}$-I orbifold model}

\paragraph{(a)}
 $v^i=1/12(1,-5,4)$ 
\hskip 1cm
$F_4 \times SU(3)$ lattice

\paragraph{(b)} Twist
\begin{eqnarray*}
&\theta e_2=\sum_{b=1}^3e_b, \quad \theta e_a=e_{a+1}, 
\quad a=1,3,5,\\
&\theta e_4=-2e_1-2e_2-e_3-e_4, \quad 
\theta e_5=-e_5-e_6.
\end{eqnarray*}

\paragraph{(c)}Eigenstates
\begin{eqnarray*} 
&|0\rangle, \quad 
|e_3\rangle +\alpha|e_3+e_4\rangle +\alpha^2|e_4\rangle,\\
&|e_1\rangle + \beta |\sum_{a=1}^4e_a\rangle 
+\beta^2 |e_2+e_3\rangle \\
&+\beta^3|e_1+e_3+e_4\rangle + \beta^4 |e_1+e_2+e_3\rangle 
+\beta^5 |e_2\rangle,\\
&|e_1+e_2\rangle + \beta |e_2+e_3+e_4\rangle 
+\beta^2 |e_1+e_4\rangle \\
&+\beta^3|e_1+e_2+e_4\rangle + \beta^4 |e_2+e_4\rangle 
+\beta^5 |e_1+e_3\rangle,\\
\end{eqnarray*}
where $\alpha={\rm exp}[2\pi im/3]$ and $m=0,1,2$ and 
$\beta={\rm exp}[\pi ik/3]$ and $k=0,1,\cdots,5$.

\paragraph{(d)}Wilson lines \ 
$3a_5 \approx 0$, \ 
($a_5 \approx a_6$, \ $a_b \approx 0$ \ ($b=1\sim4$)).

\subsection*{A.7 $Z_{12}$-II orbifold model}

\paragraph{(a)}
 $v^i=1/12(1,5,-6)$ 
\hskip 1cm
$F_4 \times SU(2)^2$ lattice

\paragraph{(b)} Twist
\begin{eqnarray*}
\theta e_a=-e_a, \quad a=5,6.
\end{eqnarray*}
The others $e_a$ ($a=1 \sim 4$) are transformed in the same way as 
those for $T_6$ of the $Z_{12}$-I orbifold models.

\paragraph{(c)}Eigenstates are same those of $T_4$ for the $Z_{12}$-I 
orbifold models.

\paragraph{(d)}Wilson lines \
$2a_5 \approx 2a_6 \approx 0$, \ 
($a_b \approx 0$ \ ($b=1\sim4$)).

\newpage
\pagestyle{empty}
\section*{Table Captions}

\renewcommand{\labelenumi}{Table~\arabic{enumi}}
\begin{enumerate}
\item Conjugation of $\tilde \Delta_n$ and degeneracy factors.
The second column shows the conjugation of $\tilde \Delta_n$
under $P+nV$ $\to$ $-(P+nV)$.
In the third column, the common number of degeneracy factors are shown
for $\tilde \Delta_n = \omega_{(n)}^k$ and $\tilde \Delta_n 
= \omega_{(n)}^{-k-\delta}$.

\item Shifts for $Z_4$ orbifold models.
The second column shows the independent $E_8$ shifts.
The third column shows the gauge group.
In the fourth and fifth columns, untwisted matters are shown in
$U_1$ $\&$ $U_2$ and $U_3$ subsectors, respectively.

\item Shifts for $Z_6$ orbifold models.
The second column shows the independent $E_8$ shifts.
The third column shows the gauge group.
In the fourth, fifth and sixth columns, untwisted matters are shown in
$6PV=1$, $6PV=2$ and $6PV=3$ subsectors, respectively.

\item $Z_4$ orbifold models.
The second column shows gauge groups omitting $U(1)$ groups.
The third column shows combinations of $E_8$ shifts and $E'_8$ shifts 
as the corresponding numbers of the first column in Table 2.
Massless states of $T_1$ and $T_2$ are shown in the fourth and fifth 
columns.

\item $T_3$ of $Z_6$-I orbifold models.
The third column shows combinations of $E_8$ shifts and $E'_8$ shifts 
as the corresponding numbers of the first column in Table 3.
Massless states of $T_3$ are shown in the fourth column.
In the column of gauge group, $G_{6,3,2}$ denotes 
$SU(6)\times SU(3) \times SU(2)$ and $U(1)$ groups are omitted.

\item $T_3$ of $Z_6$-II orbifold models.
The third column shows combinations of $E_8$ shifts and $E'_8$ shifts 
as the corresponding numbers of the first column in Table 3.
Massless states of $T_3$ are shown in the fourth column.
In the column of gauge group, $G_{6,3,2}$ denotes 
$SU(6)\times SU(3) \times SU(2)$ and $U(1)$ groups are omitted.

\end{enumerate}

\newpage
\begin{center}
Table 1. conjugation of $\tilde \Delta_n$ and degeneracy factors.
\end{center}

\begin{center}
\begin{tabular}{|c|c|c|}\hline
Orbifold & conjugation of $\tilde \Delta_n$ & degeneracy \\ \hline 
\hline
$Z_4$ & $\tilde \Delta_2=1 \longleftrightarrow \tilde \Delta_2=-1$ 
& 6 \\ \hline
$Z_6$-I & $\tilde \Delta_3=e^{-2\pi i/3}$ \ invariant 
& 5 \\ 
 & $\tilde \Delta_3=1 \longleftrightarrow \Delta_3=e^{2\pi i/3}$
& 5 \\ \hline
$Z_6$-II & $\tilde \Delta_3=e^{2\pi i/3}$ \ invariant 
& 4 \\ 
 & $\tilde \Delta_3=1 \longleftrightarrow \Delta_3=e^{-2\pi i/3}$
& 4 \\ \hline
$Z_8$-I & $\tilde \Delta_4=1 \longleftrightarrow \tilde \Delta_4=i $ 
& 3 \\ 
 & $\tilde \Delta_4=-1 \longleftrightarrow \Delta_4=-i$
& 3 \\ \hline
$Z_8$-II & $\tilde \Delta_4=1 \longleftrightarrow \Delta_4=-1$  
& 4 \\ 
 & $\tilde \Delta_4=\pm i$ \ invariant 
& 3 \\ \hline
$Z_{12}$-I & $\tilde \Delta_6=1 \longleftrightarrow 
\tilde \Delta_6=e^{2\pi i/3}  $ & 3 \\ 
 & $\tilde \Delta_6 =e^{-2\pi i/3}$ \ invariant & 3 \\ 
 & $\tilde \Delta_6=-1 \longleftrightarrow 
\Delta_6=e^{-\pi i/3}$& 2 \\ 
 & $\tilde \Delta_6 =e^{\pi i/3}$ \ invariant & 2 \\ \hline
$Z_{12}$-II & $\tilde \Delta_6=e^{m\pi i/3} \longleftrightarrow 
\tilde \Delta_6=-e^{m\pi i/3} $ & 2  \\ \hline
\end{tabular}
\end{center}

\vskip 2cm 
\begin{center}
Table 2. Shifts for $Z_4$ orbifold models
\end{center}
\begin{tabular}{|c|c|c|c|c|}\hline
\#         & Shift ($4V^I$) & Gauge group & $U_1,U_2$ & $U_3$  \\
\hline \hline
0  & (00000000) & $E_8$ & & \\ \hline
1  & (22000000) & $E_7 \cdot SU_2$ & & (56,2) \\ \hline
2  & (11000000) & $E_7 \cdot U_1$ & (56) & 2(1) \\ \hline
3  & (21100000) & $E_6 \cdot SU_2\cdot U_1$ & 
$(\overline{27},2) +(1,2)$ & $(\overline{27},1) + (27,1)$ \\ \hline
4  & (40000000) & $SO_{16}$ & & $(128_s)$ \\ \hline
5  & (20000000) & $SO_{14}\cdot U_1 $ & $(64_s)$ & $2(14_v)$ \\ \hline
6  & (31000000) & $SO_{12} \cdot SU_2\cdot U_1$ & 
$(32_c,1) +(12_v,2)$ & $(32_c,1) + 2(1,1)$ \\ \hline
7  & (22200000) & $SO_{10} \cdot SU_4$ & 
$(16_c,4)$ & $(10_v,6)$ \\ \hline
8  & (31111100) & $SU_8 \cdot SU_2$ & 
$(28,2)$ & $(70,1)$ \\ \hline
9  & (1111111-1) & $SU_8 \cdot U_1$ & 
$(\overline{56})+(8)$ & $(28)+(\overline{28})$ \\ \hline
\end{tabular}
\newpage
\begin{center}
Table 3. Shifts for $Z_6$ orbifold models
\end{center}
\hskip -2cm
\begin{tabular}{|c|c|c|c|c|c|}\hline
\#         & Shift ($4V^I$) & Gauge group & $U(6PV=1)$ & $U(6PV=2)$ 
& $U(6PV=3)$  \\
\hline \hline
0  & (00000000) & $E_8$ & & & \\ \hline
1  & (33000000) & $E_7 \cdot SU_2$ & & & (56,2) \\ \hline
2  & (11000000) & $E_7 \cdot U_1$ & 56 & 1 & \\ \hline
3  & (22000000) & $E_7 \cdot U_1$ &  & 56+1 & \\ \hline
4  & (42200000) & $E_6 \cdot SU_3$ &  & 
$(\overline{27},\overline 3)$ & \\ \hline
5  & (21100000) & $E_6 \cdot SU_2 \cdot U_1$ 
& $(\overline{27},2)$ & $(27,1)$ & $2(1,2)$\\ \hline
6  & (32100000) & $E_6 \cdot U_1^2$ 
& $\overline{27}+1+1$ & $\overline{27}+1$ & $27+\overline{27}$\\ \hline
7  & (60000000) & $SO_{16}$ & & & $128_s$ \\ \hline
8  & (20000000) & $SO_{14} \cdot U_1$ & $64_c$ & $14_v$ & \\ \hline
9  & (40000000) & $SO_{14} \cdot U_1$ & & $64_c+14_v$ & \\ \hline
10 & (42000000) & $SO_{12} \cdot SU_2 \cdot U_1$ 
& $(32_s,1)$ & $(12_v,2)+(1,1)$ & $(32_c,2)$\\ \hline
11 & (51000000) & $SO_{12} \cdot SU_2 \cdot U_1$ 
& $(12_v,2)$ & $(32_s,1)+(1,1)$ & $(32_c,2)$\\ \hline
12 & (31000000) & $SO_{12} \cdot U_1^2$ 
& $32_s+12_v$ & $32_c+1+1$ & $12_v+12_v$\\ \hline
13 & (22200000) & $SO_{10} \cdot SU_3 \cdot U_1$ 
& $(16_s,\overline{3})$ & $(10_v,3)+(1,3)$ & $(16_s,1)+(16_c,1)$\\ \hline
14 & (33200000) & $SO_{10} \cdot SU^2_2 \cdot U_1$ 
& $(16_s,1,2)$ & $(16_s,2,1)$ & $(10_v,2,2)$\\ 
& & & $+(1,2,2)$ & $+(10_v,1,1)$ & \\ \hline
15 & (41100000) & $SO_{10} \cdot SU_2   \cdot U_1^2$ 
& $(16_c,1)+(10_v,2)$ & $(16_s,2)+(10_v,1)$ & $(16_c,2)+(16_s,1)$\\ 
& & & $+(1,2)$ & $+(1,1)$ & +2(1,2) \\ \hline
16 & (51110000) & $SO_8 \cdot SU_4 \cdot U_1$ 
& $(8_c,4)+(8_v,1)$ & $(8_s,4)+(1,6)$ & $(8_v,6)$\\ \hline
17 & (51111111) & $SU_9$ & & 84 & \\ \hline
18 & (1111111-1) & $SU_8 \cdot U_1$ 
& $\overline{56}$ & 28 & $8+\overline 8$\\ \hline
19 & (5111111-1) & $SU_8 \cdot U_1$ 
& $\overline{28}+1$ & $\overline{28}$ & 70 \\ \hline
20 & (7111111-1)/2 & $SU_7 \cdot SU_2 \cdot U_1$ 
& $(21,2)$ & $(\overline{35},1)+(\overline 7,1)$ & $(7,2)+(\overline 7,2)$\\ 
\hline
21 & (31111111) & $SU_7 \cdot U_1^2$ 
& $35+\overline 7+1$ & $21+\overline 7 +\overline 7$ & $21+\overline{21}$\\ 
\hline
22 & (91111111)/2 & $SU_7 \cdot U_1^2$ 
& $21+7+\overline 7$ & $35+7+1$ & $21+\overline{21}$\\ 
\hline
23 & (51111100) & $SU_6 \cdot SU_3 \cdot SU_2$ 
& $(\overline 6,\overline 3,2)$ & $(\overline {15},3,1)$ & $(20,1,2)$\\ 
\hline
24 & (93311111)/2 & $SU_6 \cdot SU_3 \cdot U_1$ 
& $(20,1)+(\overline 6,\overline3)$ & $(15,\overline 3)+(1,1)$ 
& $(6,3)+(\overline 6, \overline 3)$\\ \hline
25 & (3311111-1) & $SU_6 \cdot SU_2^2 \cdot U_1$ 
& $(\overline {15},1,2)$ & $(15,1,1)$ & $(20,2,1)$\\
& & & $+(6,2,1)$ & $+(\overline 6,2,2)$ & $+2(1,1,2)$ \\
\hline
26 & (22222000) & $SU_5 \cdot SU_4 \cdot U_1$ 
& $(\overline {10},4)+(1,\overline 4)$ & $(\overline {10},1)+(5,6)$ 
& $(5,4)+(\overline 5,\overline 4)$\\ 
\hline
\end{tabular}

\newpage
\begin{center}
Table 4. $Z_4$ orbifold models
\end{center}
\hskip -2cm
\begin{tabular}{|c|c|c|c|c|}\hline
No. & Gauge group & $V^I$ & $T_1$ & $T_2$ \\ \hline \hline 
1 & $E_6 \cdot SU_2 \cdot E'_8$ & 3;0 & 
$16(\overline{27},1,1)+32(1,2;1)$ & $10(\overline{27},1;1)+6(27,1;1)$ \\
 & & & $+80(1,1;1)$ & $+32(1,2;1)+16(1,1;1)$\\ \hline
2 & $E_6\cdot SU_2\cdot E'_7 \cdot SU'_2$ & 3;1 & 
$16(1,2;1,2)+32(1,2;1,2)$ & $10(27,1;1,1)+6(\overline{27},1;1,1)$ \\
 & & & & $+32(1,2;1,1)+16(1,1;1,1)$\\ \hline
3 & $E_6\cdot SU_2\cdot SO'_{16}$ & 3;4 & 
$16(1,1;16_v)$ & $10(\overline{27},1;1)+6(27,1;1)$ \\
 & & & & $+32(1,2;1)+16(1,1;1)$\\ \hline
4 & $SO_{14}\cdot E'_7$ & 5;2 & 
$16(14_v;1)+96(1;1)$ & $16(14_v;1)+32(1;1)$ \\ \hline
5 & $SO_{14}\cdot SO'_{12} \cdot SU'_2$ & 5;6 & 
$16(1;12_v,1)+32(1;1,2)$ & $16(14_v;1,1)+32(1;1,1)$ \\ \hline
6 & $SO_{10}\cdot SU_4 \cdot E'_7$ & 7;2 & 
$16(16_s,1;1)+32(1,4;1)$ & $16(10_v,1;1)+16(1,6;1)$ \\ \hline
7 & $SO_{10}\cdot SU_4 \cdot SO'_{12}\cdot SU'_2$ & 7;6 & 
$16(1,4;1,2)$ & $16(10_v,1;1,1)+16(1,6;1,1)$ \\ \hline
8 & $SU_8 \cdot SU_2\cdot E'_8$ & 8;0 & 
$16(8,2;1)+32(\overline 8,1,;1)$ & $10(\overline{28},1;1)+6(28,1;1)$ \\
 & & & & $+32(1,2;1)$\\ \hline
9 & $SU_8 \cdot SU_2\cdot E'_7 \cdot SU'_2$ & 8;1 & 
$16(\overline 8,1;1,2)$ & $10(28,1;1,1)+6(\overline{28},1;1,1)$ \\
 & & & & $+32(1,2;1,1)$\\ \hline
10 & $SU_8 \cdot SU_2\cdot SO'_{16}$ & 8;4 & 
 & $10(\overline{28},1;1)+6(28,1;1)$ \\
 & & & & $+32(1,2;1)$\\ \hline
11 & $SU_8 \cdot E'_6 \cdot SU'_2$ & 9;3 & 
$16(8;1,1)+16(1;1,2)$ & $10(\overline 8;1,2)+6(8;1,2)$ \\
 & & & $+32(1;1,1)$ &\\ \hline
12 & $SU_8 \cdot SU'_8 \cdot SU'_2$ & 9;8 & 
$16(1;\overline 8,1)$ & $10(8;1,2)+6(\overline 8;1,2)$ \\ \hline
\end{tabular}

\newpage
\begin{center}
Table 5-1. $T_3$ of $Z_6$-I orbifold models
\end{center}

\hskip -2cm
\begin{tabular}{|c|c|c|c|}\hline
No. & Gauge group & $V^I$ & $T_3$  \\ \hline \hline 
1 & $E_7\cdot SU_2\cdot E'_8$ & 1;0 & 
$5(56,1;1)+22(1,2;1)$ \\ \hline
2 & $E_7\cdot SU_2\cdot E'_6 \cdot SU'_3$ & 1;4 & 
$5(56,1;1,1)+22(1,2;1,1)$ \\ \hline
3 & $E_6\cdot SU_2\cdot E'_8 $ & 5;0 & 
$5(27,1;1)+6(\overline{27},1;1)+22(1,2;1)+10(1,1;1)$ \\ \hline
4 & $E_6\cdot SU_2\cdot E'_6 \cdot SU'_3$ & 5;4 & 
$5(27,1;1,1)+6(\overline{27},1;1,1)+22(1,2;1,1)+10(1,1;1,1)$ \\ \hline
5 & $SO_{16}\cdot E'_7 \cdot SU'_2$ & 7;1 & 
$5(16_v;1,2)$ \\ \hline
6 & $SO_{16}\cdot E'_6 \cdot SU'_2$ & 7;5 & 
$5(16_v;1,2)$ \\ \hline
7 & $SO_{14}\cdot E'_7 $ & 8;2 & 
$11(14_v;1)+21(1;1)$ \\ \hline
8 & $SO_{14}\cdot E'_6 $ & 8;6 & 
$11(14_v;1)+21(1;1)$ \\ \hline
9 & $E_7\cdot SO'_{14} $ & 2;9 & 
$5(56;1)+42(1;1)$ \\ \hline
10 & $E_6\cdot SO'_{14} $ & 6;9 & 
$6(27;1)+5(\overline{27};1)+52(1;1)$ \\ \hline
11 & $SO_{14}\cdot SO'_{12}\cdot SU'_2 $ & 8;11 & 
$5(14_v;1,2)+11(1;1,2)$ \\ \hline
12 & $SO_{12}\cdot SU_2 \cdot SO'_{14}$ & 11;9 & 
$5(32_c,1;1)+11(12_v,1;1)+22(1,2;1)$ \\ \hline
13 & $SO_{12} \cdot E'_7$ & 12;3 & 
$5(32_s;1)+11(12_v;1)+42(1;1)$ \\ \hline
14 & $SO_{12}\cdot SU_2 \cdot SO'_{12}$ & 10;12 & 
$11(12_v,1;1)+21(1,2;1)$ \\ \hline
15 & $SO_{10}\cdot SU_3 \cdot E'_7 \cdot SU'_2$ & 13;1 & 
$5(10_v,1;1,2)+5(1,3;1,2)+6(1,\overline 3;1,2)$ \\ \hline
16 & $SO_{10}\cdot SU_3 \cdot E'_6 \cdot SU'_2$ & 13;5 & 
$5(10_v,1;1,2)+5(1,3;1,2)+6(1,\overline 3;1,2)$ \\ \hline
17 & $SO_{10}\cdot SU_2^2 \cdot E'_7 $ & 14;3 & 
$6(16_s,1,1;1)+5(16_c,1,1;1)+5(10_v,2,1;1)$ \\
& & & $+11(1,2,1;1)+22(1,1,2;1)$ \\ \hline
18 & $SO_{12}\cdot SU_2 \cdot SO'_{10} \cdot SU'^2_2$ & 10;14 & 
$5(12_v,1;1,1,2)+11(1,2;1,1,2)$ \\ \hline
19 & $SO_{10}\cdot SU_2 \cdot E'_8 $ & 15;0 & 
$6(16_s,1;1)+5(16_c,1;1)+5(10_v,2;1)$ \\
& & & $+11(1,2;1)+42(1,1;1)$ \\ \hline
20 & $SO_{10}\cdot SU_2 \cdot E'_6 \cdot SU'_3 $ & 15;4 & 
$6(16_s,1;1,1)+5(16_c,1;1,1)+5(10_v,2;1,1)$ \\
& & & $+11(1,2;1,1)+42(1,1;1,1)$ \\ \hline
21 & $SO_{16}\cdot SO'_{10} \cdot SU'_2$ & 7;15 & 
$11(16_v;1,1)$ \\ \hline
22 & $SO_{10}\cdot SU_3\cdot SO'_{10} \cdot SU'_2$ & 13;15 & 
$11(10_v,1;1,1)+11(1,3;1,1)+10(1,\overline 3;1,1)$ \\ \hline
23 & $SO_8 \cdot SU_4\cdot E'_7$ & 16;2 & 
$11(8_s,1;1)+10(1,4;1)+11(1,\overline 4;1)$ \\ \hline
24 & $SO_8 \cdot SU_4\cdot E'_6$ & 16;6 & 
$11(8_s,1;1)+10(1,4;1)+11(1,\overline 4;1)$ \\ \hline
25 & $SO_8 \cdot SU_4\cdot SO'_{12}\cdot SU'_2$ & 16;11 & 
$5(8_s,1;1,2)+6(1,4;1,2)+5(1,\overline 4;1,2)$ \\ \hline
26 & $SO_{12} \cdot SU'_9$ & 12;17 & 
$5(32_s;1)+11(12_v;1)+42(1;1)$ \\ \hline
27 & $SO_{10}\cdot SU_2^2 \cdot SU'_9 $ & 14;17 & 
$6(16_s,1,1;1)+5(16_c,1,1;1)+5(10_v,2,1;1)$ \\
& & & $+11(1,2,1;1)+22(1,1,2;1)$ \\ \hline
28 & $SU_8 \cdot SO'_{12}$ & 18;12 & 
$10(8;1)+11(\overline 8;1)$ \\ \hline
\end{tabular}

\newpage
\begin{center}
Table 5-2. $T_3$ of $Z_6$-I orbifold models
\end{center}
\hskip -2cm
\begin{tabular}{|c|c|c|c|}\hline
No. & Gauge group & $V^I$ & $T_3$  \\ \hline \hline 
29 & $SU_8 \cdot SO'_{10} \cdot SU'^2_2$ & 18;14 & 
$6(8;1,1,2)+5(\overline 8;1,1,2)$ \\ \hline
30 & $SU_8 \cdot SO'_{12}$ & 19;12 & 
$11(8;1)+10(\overline 8;1)$ \\ \hline
31 & $SU_8 \cdot SO'_{10} \cdot SU'^2_2$ & 19;14 & 
$5(8;1,1,2)+6(\overline 8;1,1,2)$ \\ \hline
32 & $SO_{14} \cdot SU'_7 \cdot SU'_2$ & 8;20 & 
$5(14_v;1,2)+11(1;1,2)$ \\ \hline
33 & $SU_7\cdot SU_2 \cdot SO'_{14} $ & 20;9 & 
$6(21,1;1)+5(\overline{21},1;1)+5(7,1;1)$ \\
& & & $+5(\overline 7,1;1)+22(1,2;1)$ \\ \hline
34 & $SO_8 \cdot SU_4 \cdot SU'_7 \cdot SU'_2$ & 16;20 & 
$5(8_s,1;1,2)+6(1,4;1,2)+5(1,\overline 4;1,2)$ \\ \hline
35 & $SU_7 \cdot E'_7 $ & 21;2 & 
$10(7;1)+11(\overline 7;1)+22(1;1)$ \\ \hline
36 & $SU_7 \cdot E'_6 $ & 21;6 & 
$10(7;1)+11(\overline 7;1)+22(1;1)$ \\ \hline
37 & $SU_7 \cdot SO'_{12} \cdot SU'_2 $ & 21;11 & 
$6(7;1,2)+5(\overline 7;1,2)+10(1;1,2)$ \\ \hline
38 & $SU_7 \cdot SU'_7 \cdot SU'_2 $ & 21;20 & 
$6(7;1,2)+5(\overline 7;1,2)+10(1;1,2)$ \\ \hline
39 & $SU_7 \cdot E'_7 $ & 22;3 & 
$5(21;1)+6(\overline {21};1)+5(7;1)+5(\overline 7,1)+42(1;1)$ \\ \hline
40 & $SO_{12} \cdot SU_2 \cdot SU'_7 $ & 10;22 & 
$11(12_v,1;1)+21(1,2;1)$ \\ \hline
41 & $SU_7 \cdot SU'_9 $ & 22;17 & 
$5(21;1)+6(\overline {21};1)+5(7;1)+5(\overline 7;1)+42(1;1)$ \\ \hline
42 & $SU_8 \cdot SU'_7 $ & 18;22 & 
$10(8;1)+11(\overline 8;1)$ \\ \hline
43 & $SU_8 \cdot SU'_7 $ & 19;22 & 
$11(8;1)+10(\overline 8;1)$ \\ \hline
44 & $G_{6,3,2} \cdot E'_8 $ & 23;0 & 
$5(20,1,1;1)+5(6,3,1;1)+6(\overline 6,\overline 3,1;1)+22(1,1,2;1)$ \\ 
\hline
45 & $G_{6,3,2} \cdot E'_6 \cdot SU'_3 $ & 23;4 & 
$5(20,1,1;1,1)+5(6,3,1;1,1)+6(\overline 6,\overline 3,1;1,1)+22(1,1,2;1,1)$ \\ 
\hline
46 & $SO_{16} \cdot G'_{6,3,2} $ & 7;23 & 
$5(16_v;1,1,2)$ \\ \hline
47 & $SO_{10} \cdot SU_3\cdot G'_{6,3,2}$ & 13;23 & 
$5(10_v,1;1,1,2)+5(1,3;1,1,2)+6(1,\overline 3;1,1,2)$ \\ \hline
48 & $SO_{14}\cdot SU'_6 \cdot SU'_3$ & 8;24 & 
$11(14_v;1,1)+21(1;1,1)$ \\ \hline
49 & $SU_6 \cdot SU_3 \cdot SO'_{14}$ & 24;9 & 
$5(20,1;1)+6(6,3;1)+5(\overline 6,\overline 3;1) +42(1,1;1)$ \\ \hline
50 & $SO_8 \cdot SU_4 \cdot SU'_6 \cdot SU'_3$ & 16;24 & 
$11(8_s,1;1,1)+10(1,4;1,1)+11(1,\overline 4;1,1)$ \\ \hline
51 & $SU_7 \cdot SU'_6 \cdot SU'_3$ & 21;24 & 
$10(7;1,1)+11(\overline 7;1,1)+22(1;1,1)$ \\ \hline
52 & $SU_6 \cdot SU^2_2 \cdot E'_7 \cdot SU'_2$ & 25;1 & 
$6(6,1,1;1,2)+5(\overline 6,1,1;1,2)+5(1,2,2;1,2)$ \\ \hline
53 & $SU_6 \cdot SU^2_2 \cdot E'_6 \cdot SU'_2$ & 25;5 & 
$6(6,1,1;1,2)+5(\overline 6,1,1;1,2)+5(1,2,2;1,2)$ \\ \hline
54 & $SU_6 \cdot SU^2_2 \cdot SO'_{10} \cdot SU'_2$ & 25;15 & 
$10(6,1,1;1,1)+11(\overline 6,1,1;1,1)+11(1,2,2;1,1)$ \\ \hline
55 & $SU_6 \cdot SU^2_2 \cdot G'_{6,3,2}$ & 25;23 & 
$6(6,1,1;1,1,2)+5(\overline 6,1,1;1,1,2)+5(1,2,2;1,1,2)$ \\ \hline
56 & $SU_5 \cdot SU_4 \cdot SO'_{12} $ & 26;12 & 
$11(5,1;1)+10(\overline 5,1;1)+11(1,6;1)$ \\ \hline
57 & $SU_5 \cdot SU_4 \cdot SO'_{10} \cdot SU'^2_2$ & 26;14 & 
$5(5,1;1,1,2)+6(\overline 5,1;1,1,2)+5(1,6;1,1,2)$ \\ \hline
58 & $SU_5 \cdot SU_4 \cdot SU'_7 $ & 26;22 & 
$11(5,1;1)+10(\overline 5,1;1)+11(1,6;1)$ \\ \hline
\end{tabular}

\newpage
\begin{center}
Table 6-1. $T_3$ of $Z_6$-II orbifold models
\end{center}
\hskip -2cm
\begin{tabular}{|c|c|c|c|}\hline
No. & Gauge group & $V^I$ & $T_3$  \\ \hline \hline 
1 & $E_7\cdot E'_8$ & 2;0 & 
$4(56;1)+44(1;1)$ \\ \hline
2 & $E_7\cdot SU_2 \cdot E'_7$ & 1;3 & 
$4(56,1;1)+20(1,2;1)$ \\ \hline
3 & $E_7 \cdot E'_6 \cdot SU'_3$ & 2;4 & 
$4(56,1;1)+44(1,1;1)$ \\ \hline
4 & $E_6 \cdot SU_2 \cdot E'_7 $ & 5;3 & 
$8(27,1;1)+4(\overline {27},1;1)+20(1,2;1)+8(1,1;1)$ \\ \hline
5 & $E_6 \cdot E'_8 $ & 6;0 & 
$4(27;1)+8(\overline {27};1)+52(1;1)$ \\ \hline
6 & $E_6 \cdot E'_6 \cdot SU'_3$ & 6;4 & 
$4(27;1,1)+8(\overline {27};1,1)+52(1;1,1)$ \\ \hline
7 & $SO_{16} \cdot E'_7 $ & 7;2 & $12(16_v;1)$ \\ \hline
8 & $SO_{16} \cdot E'_6 $ & 7;6 & $12(16_v;1)$ \\ \hline
9 & $SO_{12} \cdot SU_2 \cdot E'_7 \cdot SU'_2$ & 10;1 
& $4(12_v,1;1,2)+12(1,2;1,2)$ \\ \hline
10 & $SO_{12} \cdot SU_2 \cdot E'_6 \cdot SU'_2$ & 10;5 
& $4(12_v,1;1,2)+12(1,2;1,2)$ \\ \hline
11 & $SO_{12} \cdot SU_2 \cdot E'_8 $ & 11;0 
& $4(32_c,1;1)+12(12_v,1;1)+20(1,2;1)$ \\ \hline
12 & $SO_{12} \cdot SU_2 \cdot E'_6 \cdot SU'_3$ & 11;4 
& $4(32_c,1;1,1)+12(12_v,1;1,1)+20(1,2;1,1)$ \\ \hline
13 & $SO_{16} \cdot SO_{12} \cdot SU'_2$ & 7;11 
& $4(16_v;1,2)$ \\ \hline
14 & $SO_{14} \cdot SO'_{12}$ & 8;12 
& $12(14_v;1)+20(1;1)$ \\ \hline
15 & $SO_{12} \cdot SO'_{14}$ & 12;9 
& $4(32_s;1)+12(12_v;1)+44(1;1)$ \\ \hline
16 & $SO_{10} \cdot SU_3 \cdot E'_7$ & 13;2 
& $12(10_v,1;1)+8(1,3;1)+12(1,\overline 3;1)$ \\ \hline
17 & $SO_{10} \cdot SU_3 \cdot E'_6$ & 13;6 
& $12(10_v,1;1)+8(1,3;1)+12(1,\overline 3;1)$ \\ \hline
18 & $SO_{10} \cdot SU_3 \cdot SO'_{12} \cdot SU'_2$ & 13;11 
& $4(10_v,1;1,2)+8(1,3;1,2)+4(1,\overline 3;1,2)$ \\ \hline
19 & $SO_{14} \cdot SO'_{10} \cdot SU'^2_2$ & 8;14 
& $4(14_v;1,1,2)+12(1;1,1,2)$ \\ \hline
20 & $SO_{10} \cdot SU^2_2 \cdot SO'_{14} $ & 14;9 
& $4(16_s,1,1;1)+8(16_c,1,1;1)+4(10_v,2,1;1)$ \\ 
 & & & $+12(1,2,1;1)+20(1,1,2;1)$ \\ \hline
21 & $SO_{10} \cdot SU_2\cdot E'_7 $ & 15;3 
& $4(16_s,1;1)+8(16_c,1;1)+4(10_v,2;1)$ \\ 
 & & & $+12(1,2;1)+44(1,1;1)$ \\ \hline
22 & $SO_{12} \cdot SU_2\cdot SO'_{10} \cdot SU'_2$ & 10;15 
& $12(12_v,1;1,1)+20(1,2;1,1)$ \\  \hline
23 & $SO_8 \cdot SU_4\cdot SO'_{12}$ & 16;12 
& $12(8_s,1;1)+12(1,4;1)+8(1,\overline 4;1)$ \\  \hline
24 & $SO_8 \cdot SU_4\cdot SO'_{10}\cdot SU'^2_2$ & 16;14 
& $4(8_s,1;1,1,2)+4(1,4;1,1,2)+8(1,\overline 4;1,1,2)$ \\  \hline
25 & $E_7 \cdot SU_2\cdot SU'_9$ & 1;17 
& $4(56,1;1)+20(1,2;1)$ \\  \hline
26 & $E_6 \cdot SU_2\cdot SU'_9$ & 5;17 
& $8(27,1;1)+4(\overline {27},1;1)+20(1,2;1)+8(1,1;1)$ \\  \hline
27 & $SO_{10} \cdot SU_2\cdot SU'_9 $ & 15;17 
& $4(16_s,1;1)+8(16_c,1;1)+4(10_v,2;1)$ \\ 
 & & & $+12(1,2;1)+44(1,1;1)$ \\ \hline
28 & $SU_8 \cdot E'_7 \cdot SU'_2 $ & 18;1 
& $4(8;1,2)+8(\overline 8;1,2)$ \\ \hline
29 & $SU_8 \cdot E'_6 \cdot SU'_2 $ & 18;5 
& $4(8;1,2)+8(\overline 8;1,2)$ \\ \hline
30 & $SU_8 \cdot SO'_{10} \cdot SU'_2 $ & 18;15 
& $12(8;1,1)+8(\overline 8;1,1)$ \\ \hline
\end{tabular}

\newpage
\begin{center}
Table 6-2. $T_3$ of $Z_6$-II orbifold models
\end{center}
\hskip -2cm
\begin{tabular}{|c|c|c|c|}\hline
No. & Gauge group & $V^I$ & $T_3$  \\ \hline \hline 
31 & $SU_8 \cdot E'_7 \cdot SU'_2 $ & 19;1 
& $8(8;1,2)+4(\overline 8;1,2)$ \\ \hline
32 & $SU_8 \cdot E'_6 \cdot SU'_2 $ & 19;5 
& $8(8;1,2)+4(\overline 8;1,2)$ \\ \hline
33 & $SU_8 \cdot SO'_{10} \cdot SU'_2 $ & 19;15 
& $8(8;1,1)+12(\overline 8;1,1)$ \\ \hline
34 & $SU_7 \cdot SU_2\cdot E'_8 $ & 20;0 
& $4(21,1;1)+8(\overline {21},1;1)+4(7,1;1)$ \\ 
 & & & $+4(\overline 7,1;1)+20(1,2;1)$ \\ \hline
35 & $SU_7 \cdot SU_2\cdot E'_6 \cdot SU'_3$ & 20;4 
& $4(21,1;1,1)+8(\overline {21},1;1,1)+4(7,1;1,1)$ \\ 
 & & & $+4(\overline 7,1;1,1)+20(1,2;1,1)$ \\ \hline
36 & $SO_{16} \cdot SU'_7 \cdot SU'_2 $ & 7;20 
& $4(16_v;1,2)$ \\ \hline
37 & $SO_{10} \cdot SU_3 \cdot SU'_7 \cdot SU'_2 $ & 13;20 
& $4(10_v,1;1,2)+8(1,3;1,2)+4(1,\overline 3;1,2)$ \\ \hline
38 & $SU_7 \cdot SO'_{12}$ & 21;12 
& $12(7;1)+8(\overline 7;1)+24(1;1)$ \\ \hline
39 & $SU_7 \cdot SO'_{10} \cdot SU'^2_2$ & 21;14 
& $4(7;1,1,2)+8(\overline 7;1,1,2)+8(1;1,1,2)$ \\ \hline
40 & $SO_{14} \cdot SU'_7$ & 8;22 
& $12(14_v;1)+20(1;1)$ \\ \hline
41 & $SU_7 \cdot SO'_{14}$ & 22;9 
& $8(21;1)+4(\overline{21};1)+4(7;1)+4(\overline 7;1)+44(1;1)$ \\ \hline
42 & $SO_8 \cdot SU_4 \cdot SU'_7$ & 16;22 
& $12(8_s,1;1)+12(1,4;1)+8(1,\overline 4;1)$ \\ \hline
43 & $SU_7 \cdot SU'_7$ & 21;22 
& $12(7;1)+8(\overline 7;1)+24(1;1)$ \\ \hline
44 & $G_{6,3,2} \cdot E'_7$ & 23;3 
& $4(20,1,1;1)+8(6,3,1;1)+4(\overline 6,\overline 3,1;1)+20(1,1,2;1)$ \\ 
\hline
45 & $SO_{12}\cdot SU_2 \cdot G'_{6,3,2}$ & 10;23 
& $4(12_v,1;1,1,2)+12(1,2;1,1,2)$ \\ \hline
46 & $G_{6,3,2} \cdot SU'_9$ & 23;17 
& $4(20,1,1;1)+8(6,3,1;1)+4(\overline 6,\overline 3,1;1)+20(1,1,2;1)$ \\ 
\hline
47 & $SU_8 \cdot G'_{6,3,2}$ & 18;23 
& $4(8;1,1,2)+8(\overline 8;1,1,2)$ \\ \hline
48 & $SU_8 \cdot G'_{6,3,2}$ & 19;23 
& $8(8;1,1,2)+4(\overline 8;1,1,2)$ \\ \hline
49 & $SU_6 \cdot SU_3 \cdot E'_8$ & 24;0 
& $4(20,1;1)+4(6,3;1)+8(\overline 6,\overline 3;1)+44(1,1;1)$ \\ 
\hline
50 & $SU_6 \cdot SU_3 \cdot E'_6 \cdot SU'_3$ & 24;4 
& $4(20,1;1,1)+4(6,3;1,1)+8(\overline 6,\overline 3;1,1)+44(1,1;1,1)$ \\ 
\hline
51 & $SO_{16} \cdot SU'_6 \cdot SU'_3$ & 7;24 
& $12(16_v;1,1)$ \\ \hline
52 & $SO_{10} \cdot SU_3 \cdot SU'_6 \cdot SU'_3$ & 13;24 
& $12(10_v,1;1,1)+8(1,3;1,1)+12(1,\overline 3;1,1)$ \\ \hline
53 & $SU_6 \cdot SU^2_2 \cdot E'_7$ & 25;2 
& $12(6,1,1;1)+8(\overline 6,1,1;1)+12(1,2,2;1)$ \\ \hline
54 & $SU_6 \cdot SU^2_2 \cdot E'_6$ & 25;6 
& $12(6,1,1;1)+8(\overline 6,1,1;1)+12(1,2,2;1)$ \\ \hline
55 & $SU_6 \cdot SU^2_2 \cdot SO'_{12} \cdot SU'_2$ & 25;11 
& $4(6,1,1;1,2)+8(\overline 6,1,1;1,2)+4(1,2,2;1,2)$ \\ \hline
56 & $SU_6 \cdot SU^2_2 \cdot SU'_7 \cdot SU'_2$ & 25;20 
& $4(6,1,1;1,2)+8(\overline 6,1,1;1,2)+4(1,2,2;1,2)$ \\ \hline
57 & $SU_6 \cdot SU^2_2 \cdot SU'_6 \cdot SU'_3$ & 25;24 
& $12(6,1,1;1,1)+8(\overline 6,1,1;1,1)+12(1,2,2;1,1)$ \\ \hline
58 & $SU_5 \cdot SU_4 \cdot E'_7 \cdot SU'_2$ & 26;1 
& $8(5,1;1,2)+4(\overline 5,1;1,2)+4(1,6;1,2)$ \\ \hline
59 & $SU_5 \cdot SU_4 \cdot E'_6 \cdot SU'_2$ & 26;5 
& $8(5,1;1,2)+4(\overline 5,1;1,2)+4(1,6;1,2)$ \\ \hline
60 & $SU_5 \cdot SU_4 \cdot SO'_{10} \cdot SU'_2$ & 26;15 
& $8(5,1;1,1)+12(\overline 5,1;1,1)+12(1,6;1,1)$ \\ \hline
61 & $SU_5 \cdot SU_4 \cdot G'_{6,3,2}$ & 26;23 
& $8(5,1;1,1,2)+4(\overline 5,1;1,1,2)+4(1,6;1,1,2)$ \\ \hline
\end{tabular}

\begin{thebibliography}{99}

\bibitem{CY}
P.~Candelas, G.~Horowitz, A.~Strominger and E.~Witten, 
Nucl. Phys. {\bf B258} (1985) 46.

\bibitem{Orbi1}
L.~Dixon, J.~Harvey, C.~Vafa and E.~Witten, Nucl. Phys. 
{\bf B261} (1985) 651; {\bf B274} (1986) 285.

\bibitem{fermi}
H.~Kawai, D.C.~Lewellen and A.N.~Shellekens, 
Phys. Rev. Lett. {\bf 57} (1986) 1853; 
Nucl. Phys. {\bf B288} (1987) 1;\\
I.~Antoniadis, C.~Bachas and C.~Kounnas, 
Nucl. Phys. {\bf B289} (1987) 87.

\bibitem{Orbi3}
Y.~Katsuki, Y.~Kawamura, T.~Kobayashi, N.~Ohtsubo, Y.~Ono
 and K.~Tanioka,
Nucl. Phys. {\bf B341} (1990) 611.

\bibitem{WL1}
L.E.~Ib\'a\~nez, H.P.~Nilles and F.~Quevedo, 
Phys. Lett. {\bf B187} (1987) 25.

\bibitem{STflat}
A.~Font, L.E.~Ib\'a\~nez, H.P.~Nilles and F.~Quevedo, Nucl.~Phys. 
{\bf B307} (1988) 109;\\
J.A.~Casas, E.K.~Katehou and C.~Mu\~noz,
 Nucl.~Phys. {\bf B317} (1989) 171.

\bibitem{STflat2}
A.~Font, L.E.~Ib\'a\~nez, H.P.~Nilles and F.~Quevedo, 
Phys.~Lett. {\bf B210} (1988) 101;\\
J.A.~Casas and C.~Mu\~noz, Phys. Lett. {\bf B209} (1988) 214;
 {\bf B214} (1988) 63.


\bibitem{SUSYdual}
N.~Seiberg and E.~Witten,
Nucl. Phys. {\bf B426} (1994) 19; Nucl. Phys. {\bf B431} (1994) 484; \\ 
N.~Seiberg, Nucl. Phys. {\bf B435} (1995) 129;\\
K.~Intriligator and N.~Seiberg, hep-th/9408155.


\bibitem{STnp}
See e.g. E.~Witten, Nucl. Phys. {\bf B443} (1995) 85;
Nucl. Phys. {\bf B460} (1996) 541;\\
N.~Seiberg and E.~Witten, RU-96-12, hep-th/9603003.


\bibitem{STflat3}
A.~Font, L.E.~Ib\'a\~nez, F.~Quevedo and A.~Sierra, 
Nucl. Phys. {\bf B331} (1990) 421.


\bibitem{Orbi2}
L.E.~Ib\'a\~nez, J.~Mas, H.P.~Nilles and F.~Quevedo, 
Nucl. Phys. {\bf B301} (1988) 157.

\bibitem{Orbi4}
T.~Kobayashi and N.~Ohtsubo,
Int. J. Mod. Phys.{\bf A9} (1994) 87.
  
\bibitem{ZNM}
A.~Font, L.E.~Ib\'a\~nez and F.~Quevedo, 
Phys. Lett. {\bf B217} (1989) 272;
T.~Kobayashi and N.~Ohtsubo,
Phys. Lett. {\bf B262} (1991) 425. 

\bibitem{Yukawa}
T.~Kobayashi and N.~Ohtsubo,
Phys. Lett. {\bf B245} (1990) 441. 

\bibitem{shift}
Y.~Katsuki, Y.~Kawamura, T.~Kobayashi, N.~Ohtsubo and K.~Tanioka,
Prog. Theor. Phys. {\bf 82} (1989) 171.

\bibitem{GSO}
I.~Senda and A.~Sugamoto, 
Nucl. Phys. {\bf B302} (1988) 291.

\bibitem{couple}
S.~Hamidi and C.~Vafa, Nucl. Phys. {\bf B279} (1987) 465;\\
L.~Dixon, D.~Friedan, E.~Martinec and S.~Shenker, 
Nucl. Phys. {\bf B282} (1987) 13.

\bibitem{nonre}
M.~Cvetic, Phys. Rev. Lett. {\bf 59} (1987) 2829;\\
A.~Font, L.E.~Ib\'a\~nez, H.P.~Nilles and F.~Quevedo, 
Phys. Lett. {\bf B213} (1988) 274;\\
T.~Kobayashi, Phys. Lett. {\bf B354} (1995) 264. 

\bibitem{fmass}
T.~Kobayashi, Phys. Lett. {\bf B358} (1995) 253. 

\bibitem{FMS}
D.~Friedan, E.~Martinec and S.~Shenker, 
Nucl. Phys. {\bf B271} (1986) 93.

\bibitem{ST-FI} 
J.~Atick, L.~Dixon and A.~Sen, Nucl.~Phys. {\bf B292} (1987) 109;\\
M.~Dine, I.~Ichinose and N.~Seiberg, Nucl.~Phys. {\bf B293} (1987) 253.


\bibitem{CP}
T.~Kobayashi and C.S.~Lim,
Phys. Lett. {\bf B343} (1995) 122.

\bibitem{Z46}
Y.~Katsuki, Y.~Kawamura, T.~Kobayashi, N.~Ohtsubo, Y.~Ono and K.~Tanioka,
Phys.~Lett. {\bf B218} (1989) 169.

\bibitem{Z812}
Y.~Katsuki, Y.~Kawamura, T.~Kobayashi, N.~Ohtsubo, Y.~Ono and K.~Tanioka,
Phys.~Lett. {\bf B227} (1989) 381.

\bibitem{Z8}
H.~Kawabe, T.~Kobayashi and N.~Ohtsubo,
Phys.~Lett. {\bf B322} (1994) 331.

\bibitem{KN}
T.~Kobayashi and H.~Nakano, preprint in preparation.


\bibitem{WL2}
T.~Kobayashi and N.~Ohtsubo,
Phys. Lett. {\bf B257} (1991) 56. 

\bibitem{GS} 
M.B.~Green and J.H.~Schwarz, Phys. Lett. {\bf B149} (1984) 117;\\
L.E.~Ib\'a\~nez, Phys. Lett. {\bf B303} (1993) 55.

\bibitem{Thres}
L.J.~Dixon, V.S.~Kaplunovsky and J.~Louis, 
Nucl.~Phys. {\bf B355} (1991) 649;\\
I.~Antoniadis, K.S.~Narain and T.R.~Taylor, 
Phys.~Lett. {\bf B267} (1991) 37.

\bibitem{DFKZ}
J.-P.~Derendinger, S.~Ferrara, C.~Kounnas and F.~Zwirner, Nucl.~Phys.
 {\bf B372} (1992) 145.

\bibitem{weight}
L.J.~Dixon, V.S.~Kaplunovsky and J.~Louis, 
Nucl.~Phys. {\bf B329} (1990) 27.

\bibitem{IL}
L.E.~Ib\'a\~nez and D.~L\"ust, Nucl. Phys. {\bf B382} (1992) 305.

\bibitem{anomcon}
T.~Kobayashi, Phys.~Lett. {\bf B326} (1994) 231;\\
H.~Kawabe, T.~Kobayashi and N.~Ohtsubo,
Nucl.~Phys. {\bf B434} (1995) 210.

\bibitem{V_DA}
M.~Dine, N.~Seiberg and E.~Witten, 
Nucl.~Phys. {\bf B289} (1987) 589.


\bibitem{ST-soft}
B.~de~Carlos, J.A.~Casas and C.~Mu\~noz,
 Phys. Lett. {\bf B299} (1993) 234;\\
V.S.~Kaplunovsky and J.~Louis, Phys. Lett. {\bf B306} (1993) 269.

\bibitem{ST-soft2}
A.~Brignole, L.E.~Ib\'a\~nez and C.~Mu\~noz,
 Nucl. Phys. {\bf B422} (1994) 125;\\
T.~Kobayashi, D.~Suematsu, K.~Yamada and Y.~Yamagishi, 
Phys. Lett. {\bf B348} (1995) 402;\\
A.~Brignole, L.E.~Ib\'a\~nez, C.~Mu\~noz and C.~Scheich, 
FTUAM 95/26, hep-ph/9508258.

\bibitem{ST-soft3}
Y.~Kawamura and T.~Kobayashi, Phys. Lett.\ {\bf B375} (1996) 141.


\bibitem{soft+flat}
Y.~Kawamura and T.~Kobayashi, in preparation.

\bibitem{BD}
T.~Banks and M.~Dine, RU-95-51, hep-th/9508071.

\end{thebibliography}
\end{document}